\documentstyle[12pt,epsf]{article}
\newcounter{nfig}
\newcommand{\figcap}{\par\refstepcounter{nfig}{Fig.\arabic{nfig}. }}
\newcommand{\cd}{\makebox[0.08cm]{$\cdot$}}
\title{Light-Front Dynamics}
\author{
V.A. Karmanov\thanks{e-mail: karmanov@sci.lebedev.ru}
\\ 
{\small \em Lebedev Physical Institute, Leninsky Prospekt 53, 117924 
Moscow, Russia} }
\begin{document}
\maketitle
\bibliographystyle{unsrt}

\abstract{The wave function in relativity is defined, in four-dimensional
space, on a space-like three-dimensional plane. The plane, most close to the 
time-like region, is the light-front plane $ct+z=0$. Corresponding dynamical 
approach -- {\em the light-front dynamics} -- has considerable advantages.  We
describe, in a field-theoretical framework, the construction of light-front
dynamics and illustrate it by some examples.}

\section{Introduction}\label{intro}

A few centures ago Galileo Galilei has discovered that the rectilinear motion
is indistinguishable from the rest. Two observers, the laboratory observer and
the moving one, carring out the same experiments, obtain the same results. 
This discovery is deeply consistent with our intuition: the observer in an
isolated laboratory does not interact with environment and, hence, he has no
any way to learn about his motion.

At the beginning of this century the existence of the limiting velocity was
established.  This is the light speed $c$. Nothing can move faster. This
discovery was also very consistent with our intuition.  Indeed, if the limiting
velocity  would not exist, a very far part of the Universe could make an
immediate inluence to us. This seems unnatural.

According to the Galilei principle, the  limiting speed should be the same in
any moving system of reference. Otherwise, the observer would be able to notice
his motion, measuring speed of light. However, this seems paradoxal from point
of view of our everyday experience. Pursuing the light, we can accelerate our 
system almost untill the light speed, but the light still runs away with the
same speed $c$.

Einstein discovered, that the Galilei principle is reconciled with existence of
the limiting velocity because of change of properties of space and time in a
moving system relative to the rest one. For both observers the space-time in
their own systems is the same, but for the observer from the rest system  the
space-time in the moving system  looks different than his own one. In
particular, when the speed $v$ of the moving system approaches to $c$, the
laboratory observer notices that the clock in this system delays from his one.
In its turn, the observer in the moving system sees the similar effect: from
his point of view the time in the rest system delays and almost stops when his
speed approaches to $c$. Not only the clock, but any physical process observed 
from the moving system is stopped as well. To describe the physical phenomena,
the laboratory observer  can use, naturally, his own clocks and the space
scales.  However, on his choice, he can use the clocks and the space scales
from the moving system. Two systems are equivalent,  but two  descriptions are
different.  The dilation of  time can be used, in a theoretical laboratory, to
make the "instant photo" of a fast, subnuclear physical process. "Stopping" the
time,  i.e., stopping the process, one obtains big advantage for study the most
fast processes proceeding with the speed close to $c$.  This dependence on the
choice of the reference frame, is, in other words, the dependence on the choice
of the space-time coordinates. In different coordinates the dynamical
description of a system is different. We get in this way the different forms of
dynamics. 

One of this form, the light-front dynamics (LFD), is very efficient tool to
investigate the field theory and, in this framework, the relativistic composite
systems (hadrons in the quark models, nuclei at relativistic relative nucleon
momenta). In this article we will show, how LFD is constructed, explain its
most principal properties, its relations to other approaches and give some
applications. There are also a lot of phenomenological applications of LFD.
They are beyond  the scope of the present paper.

%%%%%%%%%%%%%%%%%%%%%%%%%%%%%%%%%%%%%%%%%%%%%%%%%%%%%
\section{Forms of relativistic dynamics}\label{forms}

In his famous article \cite{dirac} Dirac analysed three forms of dynamics:  the
instant form, the point form and the front one.

From  the group-theoretical point of view,  the trasnformations of the system
of reference including the translations, rotaions and the Lorentz
transformations are forming the Poincar\'e group. Under the infinitesimal
transformation $g$ of the coordinate system with the translation parameters
$a_{\mu}$ and with the four-dimensional rotation parameters
$\varepsilon^{\nu\mu}$:
$$
x^{\mu}\rightarrow x'^{\mu}=x^{\mu}+a^{\mu}+\varepsilon^{\nu\mu}x_{\nu}
$$
the state vector $\phi$ is transformed as follows:
\begin{equation}\label{eq0}
\phi\rightarrow \phi'=U(g)\phi,
\end{equation}
where 
\begin{equation}\label{eq00}
U(g)=1+iP_{\mu}a^{\mu}+\frac{i}{2}J_{\mu\nu}\varepsilon^{\mu\nu}.
\end{equation}
Four translation generators $P_{\mu}$ are the operators of the four-momentum.
Six generators $J_{\mu\nu}$ of the rotaions and the Lorentz transformations are
the operators of the four-dimensional angular momentum.  The commutation
relations between them have the form:                           
\begin{eqnarray}\label{kt24}                                                    
[P_{\mu},P_{\nu}]&=&0\ ,
\nonumber\\                                                                                                              
\frac{1}{i}[{P}_{\mu},{J}_{\kappa\rho}]                                 
&=&g_{\mu\rho}{P}_{\kappa} -g_{\mu\kappa}{P}_{\rho}\ ,
 \nonumber\\                                                                                  
\frac{1}{i}[{J}_{\mu\nu},{J}_{\rho\gamma}] &=&g_{\mu\rho}               
{J}_{\nu\gamma} -g_{\nu\rho}{J}_{\mu\gamma}                             
+g_{\nu\gamma}{J}_{\mu\rho} -g_{\mu\gamma}{J}_{\nu\rho}.  
\end{eqnarray}               
The total  angular momentum of the system 
is determined by the Pauli-Lubansky vector:
$$
S_{\mu}=\frac{1}{2}\epsilon_{\mu\nu\rho\gamma}P^{\nu}J^{\rho\gamma}.
$$
The state vector $\phi^{J\lambda}(p)$ corresponding to a system with 
definite four-momentum $p_{\mu}$, mass $M$, total angular momentum $J$ 
and its projection $\lambda$ to the $z$-axis satisfies the following 
system of equations:
\begin{eqnarray} \label{kt25}                                                
P_{\mu}\ \phi^{J\lambda}(p)&=&p_{\mu}\ \phi^{J\lambda}(p)\ ,
 \nonumber\\                                                                                             
P^2\ \phi^{J\lambda}(p)&=&M^2\ \phi^{J\lambda}(p)\ ,                         
 \nonumber\\                                                                    
S^2\ \phi^{J\lambda}(p)&=&-M^2\ J(J+1)\ \phi^{J\lambda}(p)\ ,                 
 \nonumber\\                                                                    
S_{3}\ \phi^{J\lambda}(p)&=&M\ \lambda\phi^{J\lambda}(p)\ .                            
\end{eqnarray}  

A particular dynamical system is determined by the explicit form of these
generators, i.e., by a particular solution  of the commutaion relations
(\ref{kt24}). If these generators are expressed in terms of the particle
coordinates, we get a version of relativistic quantum mechanics with fixed
number of particles. If the denerators are expressed through the quantum
fields, we obtain a form of the quantum field theory. As soon as the generators
are known, the state vector is determined by eqs.(\ref{kt25}). For an
interacting system some Poincar\'e generators contain the interaction. Namely,
the generators changing the position of the surface, where the state vector is
defined, contain interaction. The generators, which do not change the position
of the surface, don't contain interaction and coincide with the generators of
free system. Using this property, one can classify the different forms of
dynamics.

%%%%%%%%%%%%%%%%%%%%%%%%%%%%%%%%%%%%%%%%%%%%%%%%%%%%%%%%%%%%
\subsection{Instant form}

The laboratory observer studies the physical processes in   the
four-dimensional space-time continuum described by the coordinates
$x=(t,\vec{r})$. The three-dimensional space $\vec{r}$ is a plane given by the
equation $t=const$. The observer studies the evolution of his physical system
from one plane $t=const$ to other one. The wave function $\psi(\vec{r},t)$ of a
quantum system, for a given $t$, is defined on this (three-dimensional) plane. 

This description in four-dimensional space, from one equal-time plane to other
one, corresponding to the different time instants $t=const$, is called the
instant form of dynamics. In our everyday life we always use the instant form.

The time translations of the three-dimensional plane are determined by the
Hamiltonian $H=P_0$. The interaction enters also into three operators of the
Lorentz trasnformation $J_{i0}$, $i=1,2,3$. Indeed, two simultaneous events in
one system of reference are not   simultaneous ones in a moving system.
Therefore, the Lorentz trasnformations  don't leave the plane $t=const$
invariant, they change the orientation of this plane  relative to the time
axis. This is the reason, why the corresponding generators contain  the
interaction. 

Other six generators, the translations and rotations inside the
three-dimensional space, namely, $\vec{P}$ and
$\vec{J}_i=\epsilon_{ijk}J^{jk}$  coincide with the generators of the free
system.

The instant form of dynamics is widely used for the relativistic
generalizations of the quantum mechanics.

%%%%%%%%%%%%%%%%%%%%%%%%%%%%%%%%%%%%%%%%%%%%%%%%%%%%%%%%%%%%%
\subsection{Point form}

In principle, one can define the wave function not only on the plane, but on
any space-like surface. Any two points of this surface can not be connected by
the light signal and, hence,  an event in one of these points cannot be a cause
of the other one. A convenient choice is the surface of hyperboloid,
$t^2-\vec{r}\,^2=const$. It is invariant under the Lorentz transformations.
With the state vector defined on the family of these hyperboloids, we obtain
the point form of dynamics.

In the point form the rotations and the Lorentz transformations don't change
the hyperboloid $t^2-\vec{r}\,^2=const$. Therefore all the six generators
$J_{\mu\nu}$ don't contain the interaction. Whereas, the translations are much
more complicated, and all the generators $P_{\mu}$ contain the interaction.
This means that the total momentum of a system is not the sum of the particle
momenta. This complicates the situation, inspite of the simplification of the
Lorentz boosts. 

%%%%%%%%%%%%%%%%%%%%%%%%%%%%%%%%%%%%%%%%%%%%%%%%%%%%%%%%%%%%%%%%%%%
\subsection{Front form}\label{frform}
The observer moving with the velocity $v$ along $z$-axis describes  a physical
process in his  coordinates $(t',x',y',z')$, which are related to the
laboratory ones by the Lorentz transformations: 
\begin{eqnarray}\label{eq1a}
z'&=&\frac{z-vt}{\sqrt{1-v^2/c^2}}
\nonumber\\
t'&=&\frac{t+zv/c^2}{\sqrt{1-v^2/c^2}}
\nonumber\\
x'&=&x,\quad y'=y
\end{eqnarray}
According to (\ref{eq1a}), the plane $t'=const$ in moving system corresponds to
$t+zv/c^2=const$ in the laboratory coordinates. The evolutions is considered
from one plane $t+zv/c^2=const$ to other one. Since the value of $const$ is not
yet specified, the factor  $1/\sqrt{1-v^2/c^2}$ can be absorbed by it.  For the
"null plane" we put  $t'\propto t+zv/c^2=0$. In the limiting case, when
$v\rightarrow c$, we get the plane determined by the equation $t'\propto
z_+=t+z/c=0$. The  wave function is defined on this plane. This equation
coincides with the equation for the light front $z=-ct$, moving along $-z$.
This is the reason, why  the description  in these coordinates is called the
front form of dynamics, or the light-front dynamics.

We emphasize that there are two equivalents points of view on LFD.  On the one
hand, we can study the  system in the instant form, i.e., at $t'=0$, but from
point of view of the system  of reference moving with the limiting speed
$v\rightarrow c$. This system  of reference is called the "infinite momentum
frame". One can equivalently describe the same system in the "normal",
laboratory frame, but in the light-front coordinates $(z_+,x,y,z_-)$, here
$z_+=t+z$ plays the role of the light-front "time", $z_-=t-z$ is a coordinate
in the light-front plane, and now we chose the unites with $c=1$. The first
approach is more convenient for intuition, the second one is more appropriate
for technical developments. The both differ from the instant form, $t=0$, in
the laboratory system. The both should give, in principle, the same results, 
as the instant form, but, as wee see, in more simple way.

From the group-theoretical point of view, in the front form of dynamics only
three generators $P_-,J_{1-},J_{2-}$ do not leave the light-front plane
invariant and contain the interaction. Other seven generators
$P_1,P_2,P_+,J_{12},J_{-+},J_{1+}$ and $J_{2+}$ are the free ones.

Note also that, for a free particle,  the relation between the energy and
momentum $p_0^2=\vec{p}\,^2+m^2$ can be rewritten in the light-front
coordinates as: $p_+p_- - \vec{p}_{\perp}^2=m^2$ (with
$\vec{p}_{\perp}=(p_1,p_2)$). So, the light-front energy $p_-$ of a free
particle is expressed  through the momentum as: $$
p_-=\frac{\vec{p}_{\perp}^2+m^2}{p_+}. $$ This expression does not contain any
square rooft, in contrast to the instant form.

%%%%%%%%%%%%%%%%%%%%%%%%%%%%%%%%%%%%%%%%%%%%%%%%%%%%%%%%%%%%
\subsection{Why LFD?}\label{why}

The main difficulty of the quantum field theory is the very complicated
structure of the state vector describing the particles and even the state
without any particles -- the vacuum state. The state vector is usually
described as a superposition of the bare quanta, corresponding to the
non-interating fields. If we  "switch off" the interaction between the fields,
the number of particles is conserved. As soon as we take into account the
interaction, the state vector is a superposition of the states with different
numbers of particles.

If interaction is a weak, like in the case of the quantum electrodynamics, it 
does not change the state vector too much.  Therefore,  the "dressed" electron
differs from the bare one only by small admixture of photon.

The situation is drastically different, when the interaction is strong. In this
case, the structure of the real particle is extremely complicated. For example,
the proton consists of three quarks, but these quarks are not the same quarks
that appear in the initial Lagrangian of the Quantum Chromodynamics (QCD). They
are so called the constituent quarks, which, in their turn, consist of the bare
quarks and the gluons. The state vector of the proton is a huge superposition
of the bare fields. It has not yet been calculated from the first principles of
QCD.

One should emphasize that not only the proton state, but also the state
without physical particles -- the vacuum state, from point of view of the
laboratory observer, is a complicated superposition of the bare particles, or,
in other words, of fluctuations of the bare fields. At the same time,
this description of emptiness in terms of the very complicated conglomerates
of particles, seems unnatural. It would be much better to work in the approach,
in  which the vacuum is indeed nothing but emptiness. Simplifying the vacuum
wave function, we simplify not only it, but also the wave function of
the proton and of other particles, eliminating from them, like in the vacuum
wave function, the fluctuations of fields. After that on can study the real,
physical structure of particles.

{\em The vacuum is nothing but emptiness just in the light-front dynamics.}
This is the principal advantage of this approach.

Qualitatively this can be understood from point of view of the  uncertainty
principle for energy and time. Consider the fluctuation creating three
particles from vacuum.  The fluctuation with the energy  $\Delta E
=\varepsilon_{\vec{k}_1}+\varepsilon_{\vec{k}_2}+  \varepsilon_{\vec{k}_3}$ may
occur for the time $\Delta t\approx \hbar/\Delta E$ (here 
$\varepsilon_{\vec{k}}=\sqrt{\vec{k}\,^2+m^2}$).  In the  infinite momentum
frame the momenta $\vec{k}_i$ and energies $\varepsilon_{\vec{k}_i}$ of any
particle increase,  $\Delta E$ tend to infinity. Therefore, the time of
fluctuation $\Delta t$ tends to zero. The contribution of this  fluctuation  to
the vacuum wave function disappears.  

This result is quite consistent with the mentioned above change of the
space-time properties in the moving system. Due to the time dilation, all
the physical processes are delaied, and the fluctuation has no time to occur.
This means that in the thought experiment in the infinite momentum frame we
study the particles prepared "far in advance", not spoiled by the vacuum
fluctuations. 

As already emphasized above, one can directly formulate the theory in the
light-front variables, without taking any infinite momentum frame limit. This
formulation includes the rules of the graph techniques, which allow to
calculate the amplitudes. In principle, they could contain the vertices
corresponding to vacuum fluctuations. We will see below that in LFD these
vertices do not appear. This is the quantitative manifestation of the of
disappearence of the vacuum fluctuations. {\em In LFD, the bare vacuum state,
i.e., the eigenstate of the free Hamiltonian, is also an eigenstate of full 
Hamiltonian, containing the interaction.} This property manifests itself in the
formalism of LFD.

%%%%%%%%%%%%%%%%%%%%%%%%%%%%%%%%%%%%%%%%%%%%%%%%%%%%%%%%%%%%
\subsection{LFD and relativistic quantum mechanics} 

The dynamics of a  nonrelativistic quantum system is determined by the
Schr\"odinger equation with appropriate interaction Hamiltonian. Similar
construction is developed for the relativistic quantum mechanical models. These
models are based not on the field theory, but on a  construction of
relativistic phenomenological Hamiltonians in terms of the particle
coordinates. The difference, in comparison to the nonrelativistic case, is in
the fact that in the relativistic case the interaction enters in a few
generators, so, we get a few "Hamiltonians". For example, in the front form,
the "potential" is introduced  in the generators $P_-,J_{1-},J_{2-}$. It has to
be introduced by a selfconsistent way, since the generators should satisfy the
proper commutaion relations of the Poincar\'e group. In this scheme one can fit
the phenomenological potential, for example, between two nucleons,  and then
describe the properties of two-nucleon system: the deuteron wave function, the
electromagnetic form factors, etc. The approach is also generalized to the
three-body case.  One can find the details in the review papers and books
\cite{kogsus73,ls78,lev83,desp90,kp91,coester92,gars93,keist94fb,zakopane,cdkm}. 
For the applications of the point form of dynamics to deep inelastic scattering
see the paper \cite{lev96}.

Below in this article we  concentrate on the field-theoretical approach in the
framework of LFD. Many other detailes can be also found in the above review
papers. 

%%%%%%%%%%%%%%%%%%%%%%%%%%%%%%%%%%%%%%%%%%%%%%%%%%%%%%%%%%%%
\subsection{Explicitly covariant LFD}

Together with big advantage of the simple vacuum structure, the light-front
dynamics with the light-front plane $z_+=t+z=0$ has a disadvantage: the
coordinates $x,y$ and $z$ appear in a non-symmetric way. Because of that the
theory loses the explicit relativistic and rotational covariance. For example,
in the pertubation theory, the amplitude in a given order is determined by sum
of a few  time-ordered graphs which differ from each other by the relative time
order of the interaction vertices. The sum of them is covariant, but any
particular term in this sum is not covariant. So, we deal with the theory,
which provides, in principle, the covariant final results, but not the
intermediate ones. Because of approximations, the covariance of the final
results can be also lost. 

Inspite of this inconvenience, LFD is applied in many papers to QCD, to the
hadrons in quark models and to the relativistic nuclear physics. The
applications to the light-front QCD and other references can be found, in
particular, in \cite{zakopane,leshouches,glazek}.  Note that in the paper
\cite{wilson} is was shown that the constituent quark picture with logarithmic
confinement naturally appears in weak coupling  light-front QCD. The
applications to the  relativistic composite systems (hadrons and nuclei) and
the corresponding references can be  found in the above review papers. The
rules of the graph techniques for the light-front quantum electrodynamics,
alternative to the Feynman ones, were developed in \cite{brs,brlep}. It has
been demonstrated that the light-front QED reproduces the results obtained in
the Feynman approach (such as anomalous magnetic moment of electron, etc.).

To avoid the inconvenience related to the absence of the covariance, the
explicitly covariant version of LFD has been proposed \cite{karm76} (see for
review \cite{cdkm}). In this version  the state vector is defined on the
light-front plane of the general position, given by the equation $\omega\cd
x=\omega_0 t-\vec{\omega}\cd \vec{r}=0$,  where $\omega= (\omega_0,
\vec{\omega})$ is a four-vector with $\omega^2= \omega_0^2-\vec{\omega}\,^2=0$.
This is a generalization of the standard light-front approach. The latter
corresponds to the particular value of $\omega=(1,0,0,-1)$. 

The covariance means that, for example, any four-vector can be transformed from
one system of reference to other one by a standard matrix, which depends on the
kinematical parameters only, relating two system of reference. Therefore, this
matrix is one and the same for all the four-vectors. 

The absence of the explicit relativistic covariance in the standard version of
LFD is related to the fact that the state vector depends dynamically on the
orientation of the light-front plane. As mentioned above, the corresponding
generators of these transformations contain the interaction. Rotating the
system of reference, we rotate this plane. So, in the standard approach, with
the light-front plane $t+z=0$, there is no any universal kinematical
transformation law for the light-front state vector. 

In the explicitly covariant version of LFD the kinematical transformations of
the system of reference are separated from the dynamical transformations of the
light-front plane. So, all the transformations of the reference system are
kinematical ones. This restors the explicit covariance. At the same time, the
dependence of the state vector on the orientation of the light front remains to
be dynamical. This orientation is determined by the direction of the
four-vector $\omega$.  The dependence of the state vector on the light-front
orientation is now nothing but the dependence of the four-vector $\omega$.
Therefore, the theory remains to be explicitly covariant.

In this scheme one can construct two sets of the Poincar\'e generators: ({\it
i}) The generators responsible for transformations of the state  vector under
transformations of the reference system; they are kinematical and don't contain
interaction. ({\it ii}) The generators responsible for transformations of the
state vector under translations and rotations of the light-front plane; they
are dynamical and contain interaction. The construction of these generators are
given in Appendix. Group-theoretical aspects of  the explicitly covariant LFD
are clarified in the paper \cite{coest99}.

%%%%%%%%%%%%%%%%%%%%%%%%%%%%%%%%%%%%%%%%%%%%%%%%%%%%%%%%%%%%%%%%
\section {S-matrix}\label{smat}

In the instant form,  the S-matrix $S(-\infty,t)$ gives the time evolution of
the wave function, defined  at $t=-\infty$, to the time $t$.  The S-matrix
$S(-\infty,+\infty)$ gives the scattering amplitude.  In LFD, this evolution
takes place from one light-front plane to other one, in the direction of the
light-front time.

As usual, the S-matrix is derived from the time-dependent
Schr\"odinger equation in the "interaction representation":
\begin{equation}\label{eq5}
i\frac{\partial \psi(t)}{\partial t}=H^{int}(t)\psi(t)
\end{equation}
where
\begin{equation}\label{eq6}
H^{int}(t)=\int H^{int}(\vec{x},t)d^3x
\end{equation}
is the interaction Hamiltonian, $H^{int}(x)=H^{int}(\vec{x},t)$ is the
Hamiltonian density.  We consider the example of the self-interacting  scalar
field: $H^{int}(x)=-g\varphi^3(x)$.  In the interaction representation the
field $\varphi(x)$ is the free field:
\begin{equation}\label{ft1}
\varphi(x)=\frac{1}{(2\pi)^{3/2}} \int\left[a(\vec{k})\exp(-ik\cd x) +
a^{\dagger}(\vec{k})\exp(ik\cd 
x)\right]\frac{d^3k}{\sqrt{2\varepsilon_k}}\ .
\end{equation}
$a^\dagger,a$ are the creation and annihilation operators satisfying the
commutaion relation 
$$[a(\vec{k}), a^\dagger(\vec{k})] = 
(2\pi)^3\delta^{(3)}(\vec{k}-\vec{k}\,').$$ The  S-matrix is obtained
as the formal solution of (\ref{eq5}):
\begin{equation}\label{rul1}                                                    
S=T\exp\left[-i\int H^{int}(x)d^4x\right].
\end{equation}

The T-product orders the operators in the ordinary time $t$.   The perturbation
theory is obtained by decomposing (\ref{rul1}) in series in the degrees of the
coupling constant. One may put in correspondance, to any given term, a Feynman
diagram and calculate the corresponding amplitude by the standard Feynman
rules. In this way, the Feynman propagators appear as the average value, over
the vacuum state, of the $T$-product:
$$
G(x-x')=i<0|T(\varphi(x)\varphi(x'))|0>.
$$
Its Fourier transform is just the  Feynman propagators:
$$
\frac{i}{p^2-m^2+i0}=-i\int G(x)\exp(ipx)d^4x.
$$

Another way to calculate the $S$-matrix is to develop the time-ordered
perturbation theory. For this aim, following to \cite{kadysh64,kadysh68} (see
for review \cite{kms72}), one should replace in (\ref{rul1}) the time-ordering
operator $T$ by the  explicit time ordering. Namely, one can represent
(\ref{rul1}) as: 
\begin{eqnarray} 
\label{rul11}                                                              
&&S= 1+ \sum_n\int(-i)^n H^{int}(x_1)\theta(t_1-t_2) H^{int}(x_2) \ldots 
\theta(t_{n-1}-t_n) H^{int}(x_n) d^4x_1\ldots d^4x_n . 
\nonumber\\  &&  
\label{rul1ab}                            
\end{eqnarray}                                   
In this way, the Feynman propagators are replaced by the average values of the
product of the operators $0|\varphi(x)\varphi(x')|0>$. There is no any 
$T$-product here, since it is taken into account  by the theta-functions.  In
the momentum space, with 
$$
\tilde{\varphi}(k)=\frac{1}{(2\pi)^{5/2}}\int \varphi(x)\exp(-ik\cd x)d^4x
=[a(-\vec{k})\theta(-k_0) 
+a^{\dagger}(\vec{k})\theta(k_0)]\sqrt{2\varepsilon_k}\delta(k^2-m^2) 
$$ 
thisresults in the contraction: 
\begin{equation}\label{rul7} 
\underbrace{\tilde{\varphi}(k)\tilde{\varphi}}(p) = 
\tilde{\varphi}(k)\tilde{\varphi}(p)
- :\!\tilde{\varphi}(k)\tilde{\varphi}(p)\!:\;
= \theta(p_0)\delta(p^2
-m^2)\delta^{(4)}(p+k)\ .                                 
\end{equation}                                                                                                                              
We would like to emphasize that the propagator (\ref{rul7}) contains the
delta-function $\delta(p^2-m^2)$, and therefore in the time-ordered graph
techniques {\em all  particles are always on their mass shells}. It is
convenient to replace in the following $\theta(p_0)$ in the propagator
(\ref{rul7}) by $\theta(\omega\cd p)$. This is always possible, since $p^2=m^2
> 0$.   

This method results in the so called old fashioned perturbation theory. The
amplitudes are  represented by the time-ordered graphs.  Instead of the Feynman
propagators, they contain in the denominators the differences of energies
between the initial and intermediate states. The detailed derivation for
arbitrary space-like plane is given in \cite{kadysh64,kadysh68,kms72}. Namely,
in the  paper \cite{kadysh64}  the state vector is considered as  evolving on
the family of planes  $\lambda\cd x =\sigma$, where
$\lambda=(\lambda_0,\vec{\lambda})$, $\lambda^2=1$. The old fashioned
perturbation theory is obtained from the graph techniques developed in
\cite{kadysh64,kadysh68,kms72} as  a particular case at $\lambda=(1,\vec{0})$.
The same method is applied to the case of ordering in the light-front time and
gives the amplitudes in LFD. Below namely the latter case will be considered in
detail. Here we illustrate in a simple example the difference between the
Feynman and  the usual time-ordered amplitudes. The amplitude for exchange by
the particle  in $s$ channel can be represented in two different forms. The
Feynman amplitude is:   
$$ M = \frac{g^2}{m^2-(k+p)^2}.  $$  
It corresponds to two terms in the old fashionedperturbation theory: 
\begin{equation}\label{rul13p}    
M=M_a+M_b=
\frac{g^2}{2\varepsilon_{\vec{k} +\vec{p}}  
\left[\varepsilon_{\vec{k}+\vec{p}}
-\varepsilon_{\vec{k}}-  \varepsilon_{\vec{p}}\right]}+
\frac{g^2}{2\varepsilon_{\vec{k}+\vec{p}}                        
\left[\varepsilon_{\vec{k}+\vec{p}}
+\varepsilon_{\vec{k}}+                     
\varepsilon_{\vec{p}}\right]}.                                               
\end{equation}                       
Two items in (\ref{rul13p}) correspond to two time-ordered graphs,  the second
one arises from the vacuum fluctuation. It disappears in the infinite momentum
frame (since $\Delta
E=\varepsilon_{\vec{k}}+\varepsilon_{\vec{p}}                        
+\varepsilon_{\vec{k}+\vec{p}}\rightarrow \infty$) and in the light-front
dynamics (see below).

Now consider the graph techniques, which is ordered in the light-front time. As
mentioned,  the LFD  Hamiltonian is defined on the light-front plane 
$\omega\cd x=\sigma$, $\sigma$ is the light-front time. Therefore, in the case
of the scalar fields, the integral over $d^3x$  in (\ref{eq6}) is replaced by
the  integration over the light-front plane: \begin{equation}\label{eq7}
H^{int}(\sigma)=\int H^{int}(x)\delta(\omega\cd x -\sigma)d^4x, \end{equation}
The S-matrix still has the form (\ref{rul11}), but now the T-product orders the
operators in the direction of $\omega$:
\begin{equation}\label{rul1a}                                                    
S=T_{\omega}\exp\left[-i\int H^{int}_{\omega}(x)d^4x\right] 
\end{equation}

The expression  (\ref{rul1a}) is then explicitly  represented in terms of the
light-front time  $\sigma=\omega\cd x$. Instead of (\ref{rul11}) we get:
\begin{eqnarray} 
\label{rul11a}                                                              
S&=& 1+ \sum_n\int
(-i)^n H^{int}_{\omega}(x_1)                                       
\theta\left(\omega\cd (x_1-x_2)\right) H^{int}_{\omega}(x_2) \ldots 
\theta\left(\omega\cd (x_{n-1}-x_n)\right) H^{int}_{\omega}(x_n) 
\nonumber\\                                                                     
&&\times d^4x_1\ldots d^4x_n \ . \label{rul2}                            
\end{eqnarray}                                                                  
The index $\omega$ at $H^{int}_{\omega}$ indicates that $H^{int}$ and 
$H^{int}_{\omega}$ may differ from each other in order to provide the
equivalence between (\ref{rul1}) and (\ref{rul1a}). The region where this can
happen is a line on  the light cone. Indeed, if $(x_1  -x_2)^2 >0$, the signs
of $\omega \cd (x_1 -x_2)$ and $t_1 -t_2$  are the same and hence
$H^{int}_{\omega}=H^{int}$.  If $(x_1 -x_2)^2<0$,  the operators commute:
$$                                                   
[H^{int}(x_1),H^{int}(x_2)]=0,                                                                                                                  
$$
and their relative order has no significance. On the light cone, i.e.  if $(x_1
-x_2)^2=0$,   $\omega\cd (x_1 -x_2)$ can be equal to zero while $t_1 -t_2$ may
be  different from zero.  If the integrand has no singularity at $(x_1
-x_2)^2=0$, this line does not contribute to the integral over the volume
$d^4x$. However, if the integrand is singular, some care is needed. To
eliminate the influence of this region on the $S$-matrix, we have introduced in
(\ref{rul2}) a new Hamiltonian  $H^{int}_{\omega}$, such that expressions
(\ref{rul1}) and  (\ref{rul2})  be equal to each other.  The form of
$H^{int}_{\omega}$, which provides  this  equivalence, depends on the 
singularity of the commutator  at $(x_1 -x_2)^2=0$. For  the  scalar fields,
the singularity is weak enough, and the expressions  (\ref{rul1}) and
(\ref{rul2}) are the same, so that  $H^{int}_{\omega}=H^{int}$. For fields with
spins 1/2 and 1 or with  derivative couplings, the equivalence is obtained
with  $H^{int}_{\omega}$ different from $H^{int}$ by an additional contribution
(counter term) leading  to the contact terms in the propagators (or so called
instantaneous  interaction) \cite{cdkm}.  

Introducing the Fourier transform of the Hamiltonian:        
\begin{equation}\label{rul4}                                                    
\tilde{H}_{\omega}(p)=\int H^{int}_{\omega}(x)\exp(-ip\cd x)d^4x\ ,                         
\end{equation}                                                                  
and using the integral representation for the $\theta$ function:      
\begin{equation}\label{rul5}                                                    
\theta\left(\omega\cd (x_1-x_2)\right)=  \frac{1}{2\pi i} 
\int_{-\infty}          
^{+\infty}\frac{\exp\left(i\tau\omega\cd (x_1                                      
-x_2)\right)}{\tau-i\epsilon}\ d\tau\ , \end{equation}                          
we can transform the expression (\ref{rul2}) to the form:                       
\begin{eqnarray}\label{rul6}                                                    
S&=&1-i\tilde{H}_{\omega}(0)\nonumber\\
&+&\sum_{n\geq 2} (-i)^n \int \tilde{H}_{\omega}(-\omega\tau_1)
 \frac{d\tau_1} 
{2\pi i           
(\tau_1-i\epsilon)} \tilde{H}_{\omega}(\omega\tau_1 -\omega\tau_2) \ldots                
\frac{d\tau_{n-1}}{2\pi i(\tau_{n-1}-i\epsilon)} \tilde{H}_{\omega}(\omega               
\tau_{n-1})\ . \nonumber\\                                                      
&&                                                                              
\end{eqnarray}                                                                  
The $\tau$ variable appears here as an auxiliary variable, as defined  in
eq.(\ref{rul5}); $\omega\tau$ has the dimension of a momentum.

%%%%%%%%%%%%%%%%%%%%%%%%%%%%%%%%%%%%%%%%%%%%%%%%%%%%%%%%%%%%%%%%%%%%%
%%         
\subsection{Spin 0 system}                                                      
\label{sl-graph}                                                                 

Below we still restrict ourselves by the example of the simple interaction
Hamiltonian of the form  $H=-g\varphi^3(x)$. The covariant light-front graph
technique arises when, as usual, one  represents the expression (\ref{rul6}) in
normal form.             

The four-vectors $\omega \tau_j$ in  (\ref{rul6}) are associated with a
fictitious particle -- called  spurion -- and the factors
$1/(\tau_j-i\epsilon)$ are interpreted as  the  propagator of the spurions
responsible for taking  the intermediate states off the energy shell.  This
spurion  should be  interpreted as a convenient tool in order to take into
account  off-energy shell  effects in  the covariant formulation of LFD (in the
absence of off-mass shell effects), and not as a physical particle. It  is
absent, by  definition,  in all asymptotic, on-energy shell states. We shall
show below on  simple  examples how the spurion should be used in practical
calculations.                       

The general invariant amplitude $M_{nm}$ of a transition  $m\rightarrow n$ is
related to the $S$-matrix by:  
\begin{equation}\label{rul8} 
S_{nm}=1+\frac{i(2\pi)^4\delta^{(4)}\left(\sum_{i=1}^m 
 k_i -\sum_{i=1}^n k'_i\right)}
 {\left((2\pi)^3 2\varepsilon_{k'_1}\ldots 
(2\pi)^3 2\varepsilon_{k'_n}\ (2\pi)^3 2\varepsilon_{k_1} \dots 
(2\pi)^3 2\varepsilon_{k_m}\right)^{1/2}}M_{nm}, 
\end{equation} 
where, e.g., $\varepsilon_{k_1}=\sqrt{m_1^2+\vec{k}_1^2}$.  The cross-section
of  the process $1+2\rightarrow 3+\ldots +n$ is thus expressed as:  
\begin{equation}\label{rul9}                                                    
d\sigma= \frac{(2\pi)^4}{4j\varepsilon_{k_1}\varepsilon_{k_2}} 
|M|^2\frac{d^3k_3}{(2\pi)^3 2\varepsilon_{k_3}}\cdots 
\frac{d^3k_n}{(2\pi)^3 2\varepsilon_{k_n}} \delta^{(4)}(k_1+k_2 
-k_3-\ldots-k_n)\ , 
\end{equation} 
where $j$ is the flux density of the incident particles:  
$$ 
j\varepsilon_{k_1}\varepsilon_{k_2}=\frac{1}{2}[s-(m_1+m_2)^2]^{1/2}            
[s-(m_1-m_2)^2]^{1/2},\quad s=(k_1+k_2)^2\ .                                    
$$                                                                              

\begin{figure}[htbp]
\centerline{\epsfbox{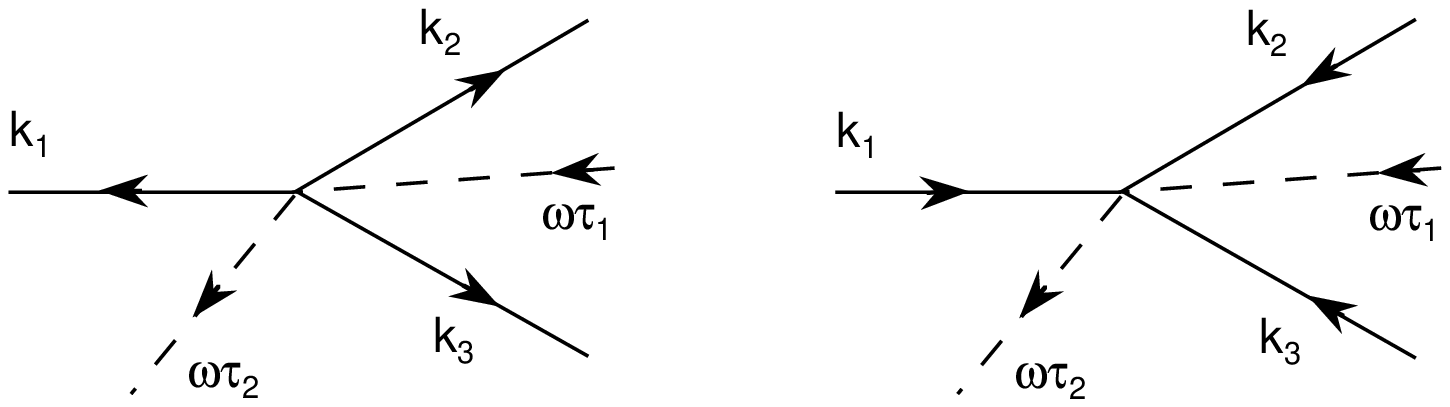}}
\figcap{The vacuum vertices.} 
\label{rulevac} 
\end{figure}

To find the matrix element $M$ of order $n$ one  must proceed as follows 
~\cite{kadysh64,kadysh68,kms72,karm76,karm88}:

\begin{enumerate}                                                               
\item \label{r1}                                                                
Arbitrary label by a number the vertices in the Feynman graph of  order  $n$. 
Orientate continuous lines (the lines of physical particles) in  the direction 
from the smaller to the larger number.  Initial particles are  oriented  as
incoming into a graph,  and final particles as  outgoing.  Connect  by  a
directed  dashed line (the spurion line) the vertices in the order of 
decreasing numbers.  Diagrams in which there are vertices with all  incoming or
outgoing particle lines (vacuum vertices, as indicated in  fig.~\ref{rulevac})
can be omitted. Associate with each continuous  line  a corresponding
four-momentum, and with each $j$-th spurion line a  four-momentum $\omega
\tau_j$.  

\item \label{r2} 
To each internal continuous line with four-momentum  $k$, associate the
propagator $\theta(\omega\cd k)\delta(k^2-m^2)$,  and  to each internal dashed
line with four-momentum $\omega\tau_j$  the  factor $1/(\tau_j-i\epsilon)$.  

\item \label{r3} 
Associate with each vertex the coupling constant $g$. All the four-momenta
at  the  vertex, {\it including the spurion momenta}, satisfy the
conservation  law, i.e., the sum of incoming momenta is equal to the sum of
outgoing  momenta. 

\item \label{r4} 
Integrate (with $d^4k/(2\pi)^3$) over those four-momenta of the  internal
particles which remain unfixed after taking into account the  conservation
laws, and over all $\tau_j$  for the spurion lines from  $-\infty$  to
$\infty$.  

\item \label{r5}  
Repeat the procedure described in \ref{r1}-\ref{r4} for all  $n!$ possible
numberings of the vertices.   
\end{enumerate}  

We omit here the factorial factors that arise from the identity of  the
particles and depend on the particular theory.  

The important property of LFD -- the disappearence of the vacuum fluctuations 
-- is just  the disappearence of the vacuum vertices  indicated in
fig.~\ref{rulevac}. In this formalism they disappear for a  trivial reason: it
is impossible to satisfy the four-momentum  conservation law for them. Indeed,
the conservation law for the vertex  of   fig. \ref{rulevac} has the form $k_1
+k_2+k_3 =  \omega(\tau_1-\tau_2)$.  Since the four-momenta are on the mass
shell: $k_{1-3}^2=m^2>0$, so that  the left-hand side is always strictly
positive: $(k_1+k_2+k_3)^2\geq  8m^2$, whereas the right-hand side is zero
since $\omega ^2=0$.  However, it will be seen that the vacuum  contributions
that vanish in the light-front approach leave their track in a different way,
making  for the fields with spin  the  light-front interaction  $H_\omega (x)$
in eq.(\ref{rul11}) different from the usual interaction $H(x)$ in
(\ref{rul1}).

The case of the particles with non-zero spins is considered in  \cite{cdkm}. In
this case, the vacuum fluctuations disappear too, but some additional (contact)
vertices appear, due to the difference between $H^{int}$ and
$H^{int}_{\omega}$. They are also taken into account by the rules of the graph
techniques.

We emphasize that despite the presence of the four-vector $\omega$ in 
eq.(\ref{rul6}), the amplitudes calculated in this way are explicitly
covariant. We just obtain the theory with separation of the kinematical
dependence of amplitudes on the reference system and of the dynamical, but
covariant dependence on the light-front orientation. The full $S$-matrix and
any physical amplitudes do not  depend on $\omega$, since eq.(\ref{rul6}) gives
the same $S$-matrix, as the initial  one, given by eq.(\ref{rul11}). However,
off-shell amplitudes  depend on  $\omega$ and off-shell light-front amplitudes
don't coincide with the Feynman ones. We will see below, that the wave
functions also depend on  $\omega$. 

The light-front diagrams can be interpreted as time-ordered graphs. As soon as
the vertices are labelled by numbers, any deformation of a diagram changing the
relative position of the vertex projections on the ``time direction" from left
to right does not change the topology of the diagram and the corresponding
amplitude. Therefore it is often convenient to deform the diagram so that the
vertices with successively increasing numbers are disposed from left to right.
This just corresponds to  time ordered graphs. In addition, this  graph
technique is three-dimensional one, i.e., the four-momenta of the particles,
even in the intermediate states, are always, on the mass shells, all the
integrations over the internal momenta are three-dimensional ones.

The light-front amplitudes can be also obtained from the graph techniques  
\cite{kadysh64,kadysh68,kms72} with $\lambda=(\lambda_0,\vec{\lambda})$, 
$\lambda^2=1$ as follows. One should replace $\lambda\rightarrow
\lambda'/\delta$  with $\lambda'^2=\delta^2$ and take limit $\delta\rightarrow
0$. This just corresponds to the infinite momentum frame limit of the
old-fashioned perturbation theory.   The light-front amplitudes can be also
obtained by direct transformation of a given Feynman amplitude
\cite{lbak,lbak2}. 

By a replacemet of variables \cite{cdkm} the covariant light-front amplitudes
can be transformed to the form of the ordinary light-front diagrams
corresponding to $\omega=(1,0,0,-1)$, given by the Weinberg rules
\cite{weinberg}.

%%%%%%%%%%%%%%%%%%%%%%%%%%%%%%%%%%%%%%%%%%%%%%%%%%%%%%%%%%%%%%%%
\subsection{Why time-ordered graphs?}

Deriving both the Feynman graph techniques and the time-ordered one, we proceed
from one and the same expression (\ref{rul1}) for the $S$-matrix and therefore
we obtain the same amplutude in a given order of the perturbation theory. The
important difference between two approaches appears in describing the bound
states, and, in general, the state vector. In the Feynman approach the bound
states are described by the  Bethe-Salpeter functions \cite{bs}, which are
defined as: 
\begin{equation} \label{bs2} 
\Phi(x_1,x_2,p)= \langle 0 \left|
T(\varphi (x_1)\varphi (x_2))\right|p\rangle .                                                                     
\end{equation}    
Here $\varphi (x)$ is the Heisenberg operator. The Bethe-Salpeter function 
depends on two four-vectors $x_{1,2}$, they include two times $t_{1,2}$. In the
momentum space the Bethe-Salpeter function looks as: $\Phi=\Phi(l_1,l_2,p)$.
Their arguments $l_{1,2}$ are off mass shell: $l_1^2\neq m^2$, $l_2^2\neq m^2$.
Though it satisfies the normalization condition, allowing to find the
normalization coefficient,  the Bethe-Salpeter function has no any
probabilistic interpretation  (see for review \cite{nak69}).

The time-ordered approach describes the bound states by means of the Fock
components. It allows to express the amplitudes in terms of the Fock componets
of the state vector.  The latters are the direct relativistic  generalization
of the non-relativistic wave functions. They depend on the on-mass-shell
four-vectors and have the same probabilistic interpretation, as the
non-relativistic wave functions. The kernel of the equation for the wave
function can be calculated by the rules of the  graph techniques. The
time-ordered graphs give also the space-time picture of the process.

The transparant physical interpretation, clear nonrelativistic limit and also
comparatively simple three-dimensional calculating formalism are the advantages
of this approach.

The relation between the light-front wave function and the Bethe-Salpeter
amplitude is given below in sect. \ref{bsf}.

%%%%%%%%%%%%%%%%%%%%%%%%%%%%%%%%%%%%%%%%%%%%%%%%%%%%%%%%%%%%%%%%        
\subsection{Simple examples}\label{simple}                                                       
                                                                                
\subsubsection{Exchange in $t$-channel}\label{tchan}                                            

\begin{figure}[hbtp]
\centerline{\epsfbox{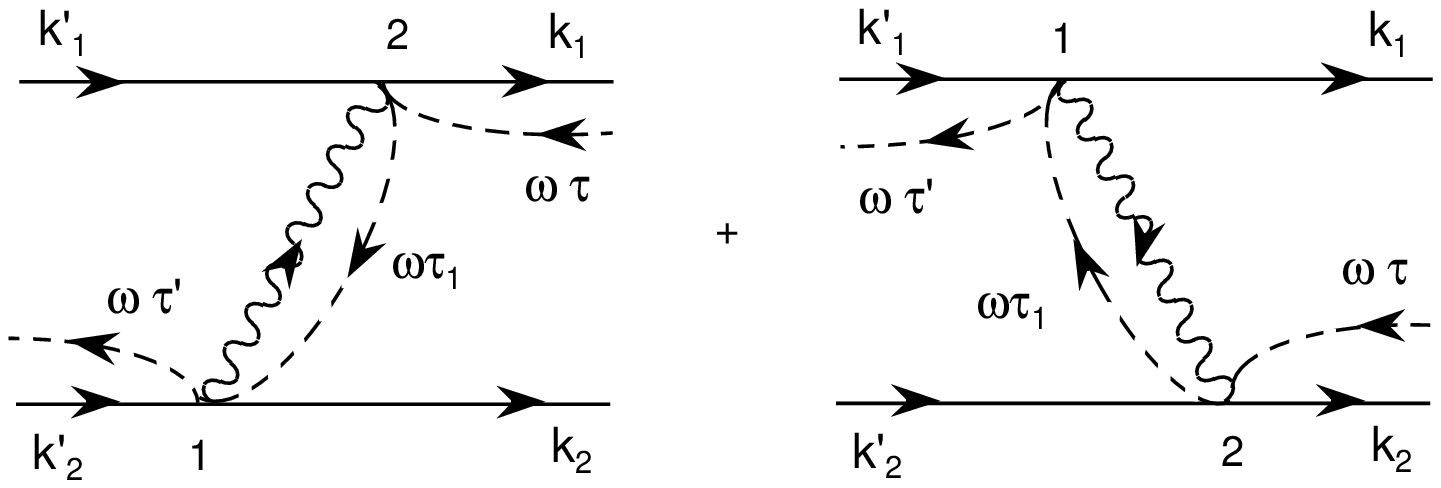}}
\figcap{Exchange by a particle in $t$-channel.}
\label{fkern} 
\end{figure}

Consider two time-ordered diagrams shown in fig.~\ref{fkern}. They correspond
to the  exchange of a scalar particle of mass $\mu$ between two scalar
particles, in  the $t$ channel. These diagrams  determine, in the  ladder
approximation, the kernel of the equation for the calculation of the 
light-front  wave function. The external spurion lines indicate that the
amplitude is off-energy shell. The term "off-energy shell", is borrowed from
the old fashioned perturbation theory, where it means that for  an amplitude
which is an internal part of a bigger diagram, there is no conservation law for
the energies of the incoming and outgoing particles (like in the  intermediate
states in the amplitudes (\ref{rul13p})). For the light-front amplitudes shown
in fig. \ref{fkern}, for $\omega=(1,0,0,-1)$, there is no  conservation law for
the minus-components of the particle momenta, i.e., for the "light-front"
energies. This momentum nonconservation is just taken into account by spurion.

According to the light-front graph technique
for spinless particles,  the amplitude has the form:   
\begin{eqnarray}\label{k1}                                                      
{\cal K}&=&\quad 
g^2\int \theta\left(\omega\cd (k_1-k'_1)\right) \delta             
\left((k_1-k'_1 +\omega\tau_1- \omega\tau)^2-\mu^2\right)                             
\frac{d\tau_1}{\tau_1-i\epsilon}
\nonumber \\                             
&&+g^2\int \theta\left(\omega\cd (k'_1-k_1)\right) 
\delta\left((k'_1-k_1 +\omega\tau_1- \omega\tau')^2-\mu^2\right)                      
\frac{d\tau_1}{\tau_1-i\epsilon}
\nonumber \\                             
&=&\quad
\frac{g^2\theta\left(\omega\cd 
(k_1-k'_1)\right)}{\mu^2-(k_1-k'_1)^2+2\tau 
\omega\cd(k_1-k'_1) -i\epsilon}
\nonumber\\ 
&&+\frac{g^2\theta\left(\omega\cd (k'_1-k_1)\right)}{\mu^2 
-(k'_1-k_1)^2 +2\tau' 
\omega\cd (k'_1-k_1)-i\epsilon}\ .  
\end{eqnarray} 
The two items in (\ref{k1}) correspond to the two diagrams of 
fig.~\ref{fkern}. They cannot be non-zero simultaneously. On the energy shell,
i.e. for both $\tau=\tau '=0$,  the expression for the kernel is identical to
the Feynman amplitude:                
\begin{equation}\label{k2}                                                      
{\cal K}(\tau=\tau'=0) = \frac{g^2}{\mu^2-(k_1-k'_1)^2-i\epsilon}\ .  
\end{equation} 
Note that the off-shell amplitude (\ref{k1}) depends on $\omega$. 

On the energy shell,  corresponding to $\tau=\tau'=0$, the dependence of the
amplitude on $\omega$  disappears. In more complicated cases, when a Feynman
diagram  corresponds  to the sum a few light-front diagrams (like in the case
of the  box  diagrams considered in sect. \ref{hfst} below), the amplitude for
a  particular light-front diagram may depend on $\omega$ even on the energy
shell. This dependence disappears in the sum of all  amplitudes in a given
order. In this case the singularities of different amplitudes, related to their
dependence on $\omega$, cancel each other in the sum.

The dependence of the perturbative amplitude (\ref{k1})  on the light-front
orientaion  (calculated exactly in the $g^2$ order) indicates that the
light-front wave function, being the off-shell object too, also depends
inevitably on the light-front orientaion (see sect. \ref{wfp} below). 

%%%%%%%%%%%%%%%%%%%%%%%%%%%%%%%%%%%%%%%%%%%%%%%%%%%%%%%%%%%%%%%%%%%%%
\subsubsection{Self-energy contributions}                         

\begin{figure}[hbtp]
\centerline{\epsfbox{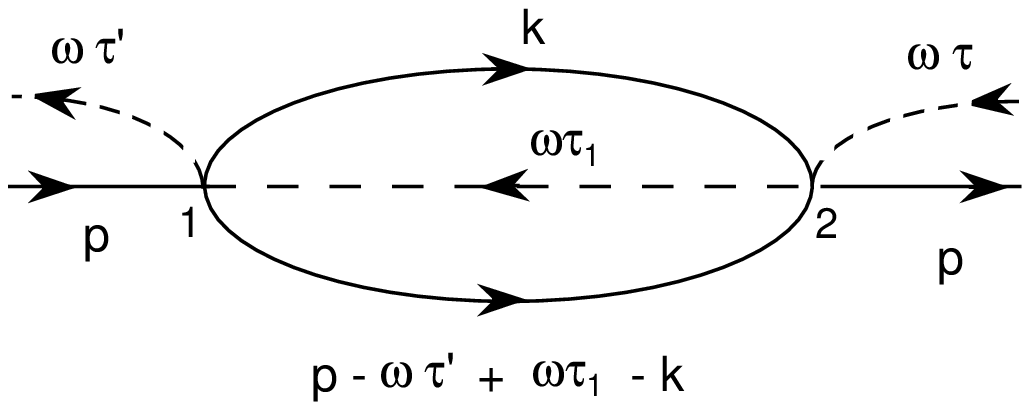}}
\figcap{Self-energy loop.}
\label{selffig} 
\end{figure}

Another simple example is the self-energy diagram shown in  fig. \ref{selffig}.
The corresponding amplitude (equal to the  self-energy up to a factor) has the
form:  
\begin{equation}\label{sel1}                                                    
\Sigma(p')=g^2\int \theta(\omega\cd k)\delta(k^2-m^2)            
\theta\left(\omega\cd (p'+\omega\tau_1-k)\right)                                    
\delta\left((p'+\omega\tau_1-k)^2 -m^2\right) \frac{d^4k}{(2\pi)^3} 
\frac{d\tau_1}{\tau_1-i\epsilon}\ ,                                                 
\end{equation}                                                                  
with $p'=p-\omega\tau'$.                                                        
                                                                                
Let $q=p'+\omega\tau_1$.  The integral over $d^4k$ is thus reduced to  the well
known calculation of the imaginary part of the Feynman amplitude, when all  the
propagators are replaced by the
delta-functions:                                                
$$                                                  
\int\delta(k^2-m^2)\delta\left((q-k)^2-m^2\right)d^4k=                         
\frac{\pi}{2\sqrt{q^2}} \sqrt{q^2-4m^2}.                                        
$$
Inserted in (\ref{sel1}), it gives:                                                      
\begin{equation}\label{sel3}                                                    
\Sigma(p') = \frac{g^2}{16\pi^2}\int\limits_{4m^2-p'^2}^{\infty} 
\frac{\sqrt{p'^2-4m^2+\tau_1}}{\sqrt{p'^2+\tau_1}}                                  
\frac{d\tau_1}{\tau_1-i\epsilon}.                                                   
\end{equation}                                                                  
The logarithmic divergence  is at the upper limit of the integration  over
$\tau_1$. One can introduce the invariant cutoff  in terms of $\tau_1$. In this
way, after renormalization, the standard expression for the self-energy
amplitude is obtained. 

The finite value of $\Sigma(p')$ for finite $\tau_1$ is  a particular
manifestation of  a general property of the light-front amplitudes. A
peculiatity of the covariant light-front amplitudes is that they have no any
ultraviolet divergences for the finite values of all the spurion four-momenta. 
All the ultraviolet divergences in all the light-front diagrams appear after
integrations over $\tau_j$ in infinite limits \cite{kadysh64}. Indeed, the
energy-momentum conservation (including the spurion  four-momentum) is valid in
any vertex.  Since all the four-momenta are on the corresponding mass  shells, 
we have at each vertex a real physical process  as far as  the kinematics is
concerned. For finite initial particle  energies and for finite incoming
spurion energy, the  energies of the particles in the intermediate states are
thus  also finite. Hence, the integrations over the particle momenta for fixed
spurion momenta are constrained by a kinematically allowed finite domain. It is
the same reason that provides  finite imaginary part of a Feynman diagram 
found by replacing the Feynman propagators
$\frac{\displaystyle{1}}{\displaystyle{(k^2-m^2+i\epsilon)}}$ by the
delta-functions $-i\pi\delta(k^2-m^2)$. In both cases the internal particle
lines are associated with the delta-functions.

The only source of the ultraviolet divergences in the light-front amplitudes is
the infinite intermediate spurion energies, i.e., infinite $\tau_j$.  This is
the  reason why divergences may appear at the upper limit of integration over
$\tau_j$. Since  $\tau_j$ are scalar quantities, one can introduce an invariant
cutoff in terms of these variables.  This way of regularizing the divergent
diagrams is another advantage of the covariant  formulation of LFD. 

For the massless particles, the light-front amplitudes may have infrared
divergences, like in the case of the Feynman diagrams.

Another peculiarity of LFD is the appearance of ``zero modes''. For
constituents of zero  mass, for instance, the state vector may contain
components with  $\omega \cd k=0$ for non-zero four-momentum $k$. In the
standard approach, this  corresponds to the finite light-front energy
$k_-=\vec{k}_{\perp}^2/k_+$ for  both $k_+=0$ and $\vec{k}_{\perp}^2=0$.  Zero
modes can also appear  in theories with spontaneously broken symmetry. They
make the equivalence between LFD and the instant form of  quantization in which
nontrivial vacuum structures (condensates) appear
\cite{zakopane,werner,heinzl,wilson94}. 

The detailed discussion of these important problems is beyond the scope of the
present paper.

%%%%%%%%%%%%%%%%%%%%%%%%%%%%%%%%%%%%%%%%%%%%%%%%%%%%%%%%%%%%%%%
\section{Light-front wave function}\label{wfp} 
As already
mentioned, the wave functions are the Fock components of the state vector
defined on the light-front plane $\omega\cd x=0$.  This means that they are
coefficients in an expansion of the state vector $|p>$ with respect to the
basis of free fields:   
\begin{eqnarray}\label{wfp1} 
|p\rangle_{\omega}&\equiv& \phi_{\omega}(p)\equiv
(2\pi)^{3/2}\int \psi_2(k_1,k_2,p,\omega\tau)
a^\dagger(\vec{k}_1)a^\dagger(\vec{k}_2)|0\rangle  \nonumber \\ &\times&
\delta^{(4)}(k_1+k_2-p-\omega\tau) 2(\omega\cd p)d\tau
\frac{d^3k_1}{(2\pi)^{3/2}\sqrt{2\varepsilon_{k_1}}}
\frac{d^3k_2}{(2\pi)^{3/2}\sqrt{2\varepsilon_{k_2}}}  + \cdots\ .  
\end{eqnarray} 
The dots $\cdots$ include the higher Fock states.  For simplicity, we omit the
spin indices.

We emphasize in (\ref{wfp1}) the presence of the delta-function 
$\delta^{(4)}(k_1+k_2 -p -\omega\tau)$. This gives the conservation law:
\begin{equation}\label{sc1}                                                     
k_1+k_2=p+\omega\tau\ .                                                         
\end{equation} 
In the particular case where $\omega =(1,0,0,-1)$, the delta-function 
$\delta^{(4)}(k_1+k_2 -p -\omega\tau)$ gives the standard  conservation  laws
for the $(\perp,+)$-components of the momenta, but does not  constrain the
minus-components.  

From (\ref{wfp1}) one can see that the wave function depends on  $\omega\tau$,
i.e., on the  orientation of the light front.  This  important property of any
Fock component is very natural. As explained above, any off-energy shell
amplitude  depends on the light-front orientation (see eq.(\ref{k1})). The
bound state wave function is always an off-shell object ($\tau\neq 0$).
Therefore it also depends on the orientation of the light-front plane. This
property is not a peculiarity of the covariant approach. At the same time, the
description of the off-energy shell effects in terms of the external spurion
lines  allows to parametrize this dependence explicitly.

%%%%%%%%%%%%%%%%%%%%%%%%%%%%%%%%%%%%%%%%%%%%%%%%%%%%%%%%%%%%%%%%%%%%%           
\subsection{The relativistic relative momentum}\label{sc}                     
We will mainly concentrate on the two-body wave function.  Generalization to
the $n$-body case is straightforward and is given in \cite{cdkm}.

\begin{figure}[htbp]
\centerline{\epsfbox{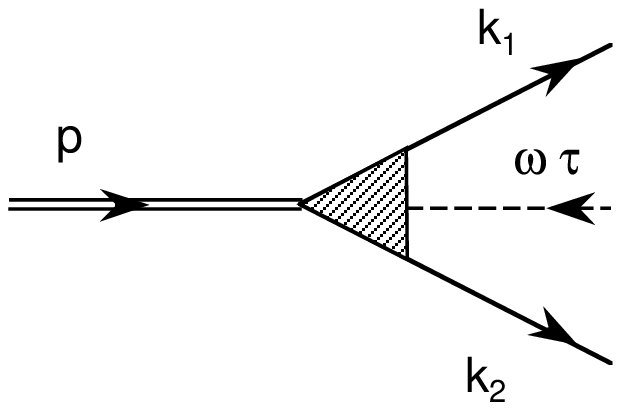}}
\figcap{Graphical representation of the two-body wave function on the
light front. The broken line corresponds to the spurion (see text).}
\label{fwf}
\end{figure}

Due to the conservation law (\ref{sc1}), the light-front wave function can be
shown graphically like a  two-body  scattering amplitude as indicated in
fig.~\ref{fwf}. The broken line  corresponds  to  the fictitious spurion. 

Due to this  analogy,  the decomposition of the wave function in independent
spin structures  and their parametrization is analogous to the expansion of a
two-body  amplitude in terms of invariant amplitudes. We will use this  analogy
below.   We emphasize again that although we assign a  momentum  $\omega\tau$
to the spurion, there is no any  fictitious particle in the physical state
vector. {\em The basis in  eq.(\ref{wfp1}) contains the particle states only}. 

The  relativistic relative momentum $\vec{k}$ has the same sense as the
norelativistic one:  it is the momentum of on of the particle  in the
c.m.-system  where   $\vec{k}_1+ \vec{k}_2=0$. Note that {\em due to the 
conservation law (\ref{sc1}), the total momentum $\vec{p} \neq 0$  of  the
system  in this reference frame is not zero}. This definition of the relative
momentum does not assume, however, that we  restrict  ourselves to this
particular reference frame.  In the arbitrary system of reference the relative
momentum is constructed by the Lorentz transformation to the system moving with
velocity
$$
\vec{v}=\vec{\cal P}/{\cal P}_0,\quad \mbox{where}\quad
{\cal P} =k_1+k_2= p + \omega\tau.
$$
We get:                                     
\begin{equation}\label{sc4}                                                     
\vec{k}= L^{-1}({\cal P})\vec{k}_1 = \vec{k}_1 -                                
\frac{\vec{\cal P}}{\cal M}\left[k_{10}                                   
- \frac{\vec{k}_1\cd\vec{{\cal P}}}{{\cal M}+{\cal P}_0}\right]\ ,              
\end{equation} 
$L^{-1}({\cal P})$ is the Lorentz boost, ${\cal M}=\sqrt{{\cal P}^2}$.                                                                 
Similarly we define the unit vector $\vec{n}$ in the direction 
of $\vec{\omega}$ in this system:
\begin{equation}\label{sc5}                                                     
\vec{n} = L^{-1}({\cal P})\vec{\omega}/|L^{-1}({\cal P})                        
\vec{\omega}| = {\cal M} L^{-1}({\cal P})                              
\vec{\omega}/\omega\cd p\ .                                                     
\end{equation}                                                                  

From these definitions, it follows that under a rotation and a  Lorentz 
transformation $g$ of the four-vectors from which $\vec{k}$ and $\vec{n}$ are
formed, the vectors $\vec{k}$ and $\vec{n}$ undergo  only          
rotations:                                                                      
$$                                                     
\vec{k}\,'=R(g,{\cal P})\ \vec{k}\ ,\quad                                       
\vec{n}'=R(g,{\cal P})\ \vec{n}\ ,                                              
$$                                                                 
where $R$ is the rotation operator:
\begin{equation}\label{rot}                                                    
R(g,p)=L^{-1}(gp)gL(p).                                        
\end{equation} 
Therefore  $\vec{k}\,^2$ and $\vec{n}\cd\vec{k}$ are the rotation and the
Lorentz invariants. For the wave function with zero angular momentum we thus
obtain~\cite{karm76}:
\begin{equation}\label{sc7}                                                     
\psi=\psi(\vec{k},\vec{n}) \equiv 
\psi(\vec{k}\,^2,\vec{n}\cd\vec{k})\ .           
\end{equation}                                                                  
It is seen from (\ref{sc7}) that the relativistic light-front wave function
depends not only on the relative momentum $\vec{k}$ but on another variable --
the unit vector $\vec{n}$.    

In the case of the states with non-zero angular momentum, the angular  momentum
is constructed by means of the spherical functions depending on the  arguments
$\vec{k}$ and $\vec{n}$.

We intro\-duce another set of vari\-ables in which the  wave  func\-tion can be
pa\-ra\-met\-rized, in analogy to the equal-time wave func\-tion in the
in\-fi\-ni\-te mo\-mentum frame. We define the           
vari\-ables:                                                                    
\begin{equation}\label{sc8}
 x=\omega\cd k_1/\omega\cd p\ , \quad R_1=k_1-xp\ ,                                                                     
\end{equation}                                                                  
and represent the spatial part of $R$ as $\vec{R}=\vec{R}_{\|} 
+\vec{R_{\perp}}$, where $\vec{R}_{\|}$ is parallel to $\vec{\omega}$  and 
$\vec{R}_{\perp}$ is orthogonal to $\vec{\omega}$.  Since 
$R\cd\omega=R_0\omega_0-\vec{R}_{\|}\cd\vec{\omega}=0$ by definition  of  $R$,
it follows that $R_0=|\vec{R}_{\|}|$, and, hence,  $\vec{R}^2_{\perp} =-R^2$ is
invariant. Therefore, $\vec{R}^2_{\perp}$  and $x$ can be chosen as
two the scalar arguments of the wave function:                                                
\begin{equation}\label{sc9}                                                     
\psi=\psi(\vec{R}^2_{\perp},x) \ .                                         
\end{equation}                                                                  
Using the definitions of the variables $\vec{R}^2_{\perp}$ and $x$, we          
can readily relate them to $\vec{k}^2$ and $\vec{n}\cd\vec{k}$:                    
\begin{equation}\label{sc9a}                                                    
\vec{R}^2_{\perp}=\vec{k}\,^2-(\vec{n}\cd\vec{k})^2,\quad                          
x=\frac{1}{2}\left(1-\frac{\vec{n}\cd\vec{k}}{\varepsilon_k}\right).               
\end{equation}                                                                  
The inverse relations are                                                       
\begin{equation}\label{sc10}                                                    
\vec{k}\,^2=                                                                    
\frac{\vec{R}^2_{\perp}+m^2}{4x(1-x)}-m^2,\quad                                 
\vec{n}\cd\vec{k}=                                                                 
\left[\frac{\vec{R}^2_{\perp}+m^2}{x(1-x)}\right]^{1/2}                         
\left( \frac{1}{2}-x\right).                                                    
\end{equation}                                                                  
The variables introduced above can  be easily generalized to the case  of
different masses and an arbitrary number of  particles~\cite{karm88}. The
corresponding variables  $\vec{q}_i,\vec{n}$ are still constructed according
to  eqs.(\ref{sc4}),  (\ref{sc5}) and the variables $\vec{R}_{i\perp},x_i$
according to (\ref{sc8}). 

%%%%%%%%%%%%%%%%%%%%%%%%%%%%%%%%%%%%%%%%%%%%%%%%%%%%%%%%%%%%%%%%%%%%%%%
\subsection{Normalization}\label{norms}
The state vector is normalized as:
\begin{equation}\label{nor1}
_{\omega}\langle p',\lambda' \vert p,\lambda\rangle_{\omega}                         
=2p_0\ \delta^{(3)}(\vec{p}- \vec{p}\,')\ \delta^{\lambda'\lambda}\ .  
\end{equation}
The Fock components are normalized so as to provide the condition 
(\ref{nor1}). Substituting the state vector (\ref{wfp1}) in the  left-hand side
of eq.(\ref{nor1}), we reproduce the right-hand side if $ \sum_n
N_n^{\lambda'\lambda} \equiv \delta^{\lambda'\lambda},   $ where
$N_n^{\lambda'\lambda}$ is the contribution to the normalization  integral from
the $n$-body Fock component. 

For the state with zero total angular momentum the normalization  condition has
the form:
\begin{equation}\label{nor4} 
\sum_n N_n=1.  
\end{equation}
In this case, the two-body contribution to the normalization integral reads:  
\begin{equation}\label{nor5} 
N_2= {1\over (2\pi)^3}\int \psi^2(\vec{k},\vec{n}) 
{d^3k\over\varepsilon_k} ={1\over (2\pi)^3}\int 
\psi^2(\vec{R}_{\perp},x) \frac{d^2R_{\perp}dx}{2x(1-x)}.
\end{equation}     

This normalization integral gives contribution only of  the two-body wave
function to the sum (\ref{nor4}). The contribution of other sectors can be
taken into account by the integral:
\begin{equation} \label{AAA}
\frac{1}{(2\pi)^3}\int\frac{d^3k}{\varepsilon_k}\frac{d^3k'}{\varepsilon_{k'}}
\psi^*(\vec{k}', \vec{n})\left[\varepsilon_k 
\delta(\vec{k}-\vec{k'})-\frac{4m^2}{(2\pi)^3}\frac{\partial 
V(\vec{k}\,',\vec{k},\vec{n},M^2)}{\partial M^2}\right] 
\psi(\vec{k}, \vec{n}) =1\ ,
\end{equation}
where $V(\vec{k}\,',\vec{k},\vec{n},M^2)$ is the  kernel of the equation for
the wave function. The second term accounts for the many-body contribution to
the norm,  $\sum_{n>2} N_n$.

%%%%%%%%%%%%%%%%%%%%%%%%%%%%%%%%%%%%%%%%%%%%%%%%%%%%%%%%%%%%%%%%%%%%%           
\subsection{Equation for the wave function} \label{toto}                                       
                                                                                
\begin{figure}[hbtp]
\centerline{\epsfbox{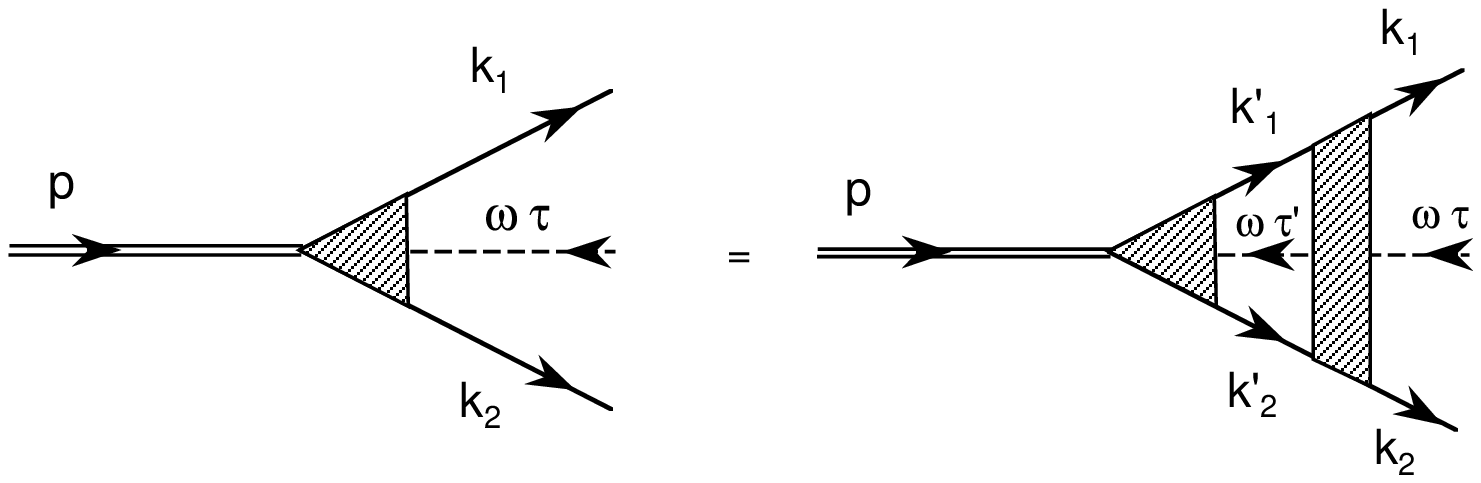}}
\figcap{Equation for the two-body wave function.}
\label{feq}
\end{figure}

The equation for the wave function is obtained  from the equation for the
vertex part shown graphically in  fig.~\ref{feq}. 

It is the analogue, for a bound state, of the  Lippmann-Schwinger equation for
a scattering state. Let us first  explain its derivation  for the case of 
spinless particles. In accordance with the rules given in 
sect.~\ref{sl-graph}, we associate with the diagram of fig.~\ref{feq}  the
following analytical
expression:                                                
\begin{eqnarray}\label{eqwf1}                                                   
\Gamma(k_1,k_2,p,\omega\tau) &=&\int \Gamma(k_1',k_2',p,\omega\tau')              
\theta(\omega \cd k_1')\delta(k_1'^2-m^2)\theta(\omega\cd k_2')                      
\delta(k_2'^2-m^2)\nonumber\\                                                   
&\times&\delta^{(4)}(k_1'+k_2'-p-\omega\tau')d^4k_1' 
{\cal K}(k_1',k_2',\omega\tau';k_1,k_2,\omega\tau)
\frac{d\tau'}{\tau'-i\epsilon}
\frac{d^4k_2'}{(2\pi)^3}\ .
\nonumber\\  
&&
\end{eqnarray} 
Here $\Gamma$ is the vertex function and the kernel ${\cal K}$ is an
irreducible block. The latter is calculated  directly by the graph technique
once the underlying dynamics is  known.   We should then express the vertex
$\Gamma$ through the two-body wave  function.  This can be done by comparing,
for example, two ways of  calculating the amplitude for the breakup of a bound
state by some  perturbation: 1) by means of the graph technique (the result
contains  $\Gamma$); 2) by calculating the matrix element of the perturbation 
operator between the bound state and the free states of $n$ particles  (the
result contains $\psi$).  We thus get:
\begin{equation}\label{eqwf2} 
\psi(k_1,k_2,p,\omega\tau)= 
\frac{\Gamma(k_1,k_2,p,\omega\tau)}{s-M^2}\ , 
\end{equation} 
where $s=(k_1+k_2)^2=(p+\omega\tau)^2$. The corresponding relation  for  the
$n$-body case has the same form. In any practical calculation of  the
amplitude, we associate $\Gamma$ with the vertex shown in  fig.~\ref{fwf} and
then express $\Gamma$ in terms of $\psi$ by  eq.(\ref{eqwf2}). 

In the simple case of a scalar particle, the equation for the wave-function in 
terms of the variables $\vec{k}, \vec{n}$ has the following form:  
\begin{equation}\label{eqwf3} 
\left(4(\vec{k}\,^2 + m^2)-M^2\right)\psi(\vec{k},\vec{n}) = 
-\frac{m^2}{2\pi^3} \int \psi(\vec{k}\,',\vec{n}) 
V(\vec{k}\,',\vec{k},\vec{n},M^2) \frac{d^3k'}{\varepsilon_{k'}} \ .  
\end{equation}                                                                  
An equation of such a type was also considered in                                  
refs.~\cite{fudaall,nam78,dn79,fn80,nw80,saw85}.                                             

In the non-relativistic limit,  equation  (\ref{eqwf3}) turns into  the
Schr\"odinger equation in momentum space, the kernel  $V$ being the
non-relativistic potential in momentum space, and the  wave  function no longer
depends on $\vec{n}$.  

We emphasize that the wave function, which is an equal-time wave function on 
the  light front, turns into the ordinary wave function in the 
non-relativistic limit where $c \to \infty$. This reflects the fact  that in
the  non-relativistic limit two simultaneous events in one frame are 
simultaneous in all other frames.  

In the variables $\vec{R}_{\perp}$ and $x$, eq.(\ref{eqwf3})  can be            
rewritten in the form:                                                          
\begin{equation}\label{eq1ab}                                                     
\left(\frac{\vec{R}^2_{\perp}+m^2}{x(1-x)}-M^2\right)                           
\psi(\vec{R}_{\perp},x)=                                                        
-\frac{m^2}{2\pi^3}\int\psi(\vec{R}'_{\perp},x')                                
V(\vec{R}'_{\perp},x';\vec{R}_{\perp},x,M^2)                                    
\frac{d^3R'_{\perp}dx'}{2x'(1-x')}\ .                                           
\end{equation}                                                                  
In this form, this equation is nothing else than the Weinberg                   
equation~\cite{weinberg}.

The advantages of the equation for the wave function in the form  (\ref{eqwf3})
compared with (\ref{eq1ab}) are its similarity to the  non-relativistic
Schr\"odinger equation in momentum space, and its  simplicity in the case of
particles with spin. These properties make  eq.(\ref{eqwf3}) very convenient
for practical calculations.  
                                                                                                                                
The kernel of eq.(\ref{eqwf3}) depends on the vector variable  $\vec{n}$. We
shall see that this dependence, especially the part which depends on $M^2$, is
associated with the  retardation of the interaction. From this point of view,
the  dependence  of the wave function $\psi(\vec{k},\vec{n})$ on $\vec{n}$ is
a  consequence of retardation. 

%%%%%%%%%%%%%%%%%%%%%%%%%%%%%%%%%%%%%%%%%%%%%%%%%%%%%%%%%%%%%%%%%%%%%    
\subsection{The Wick-Cutkosky model}                                
                     
As a simple example, we shall derive  in this  section the light-front  wave
function of a system consisting of two scalar particles with  mass  $m$
interacting through the exchange of a massless scalar particle. The kernel is
calculated  in the ladder approximation. This is the so-called  Wick-Cutkosky
model. The  diagrams that determine the kernel are shown in fig.~\ref{fkern}. 
The kernel ${\cal K}$ is given by eq.(\ref{k1}) with $\mu=0$.   Going over from
the kernel ${\cal K}$ to  $V=-{\cal  K}/(4m^2)$, introducing the constant
$\alpha=g^2/(16\pi m^2)$, and  expressing (\ref{k1}) by means of the initial
and final relative momenta $\vec{k},\vec{k}'$, we obtain~\cite{karm80}:    
\begin{equation}\label{k3}                                                      
V=-4\pi\alpha/\vec{K}^2,
\end{equation}
where                                   
\begin{equation}\label{k4}                                                      
\vec{K}\,^2 = (\vec{k}\,' - \vec{k}\,)^2 -
(\vec{n}\cd\vec{k}\,')(\vec{n}\cd\vec{k})                                           
\frac{(\varepsilon_{k'}-\varepsilon_k)^2}{\varepsilon_{k'}
\varepsilon_k} +(\varepsilon_{k'}^2
+\varepsilon_k^2-\frac{1}{2}M^2)                           
\left|\frac{\vec{n}\cd\vec{k}\,'}{\varepsilon_{k'}}                                
-\frac{\vec{n}\cd\vec{k}}{\varepsilon_k}\right|\ .                                 
\end{equation}                                                                  

For $k,k' \ll m$, eq.(\ref{k3}) turns into the Coulomb potential in momentum
space                                                    
\begin{equation}\label{k5}                                                      
V(\vec{k}\,',\vec{k})\simeq 
-\frac{4\pi\alpha}{(\vec{k}\,'-\vec{k})^2}\ . 
\end{equation}                                                                
For $\alpha\ll 1$, $|\epsilon_b|= |M-2m|=m\alpha^2/4\ll m$, the wave  function
is concentrated in the non-relativistic region of momenta.   The
non-relativistic wave function of the ground state in the Coulomb  potential
has the form:   
\begin{equation}\label{k6}                                                      
\psi(\vec{k})=\frac{8\sqrt{\pi m}\kappa^{5/2}} 
{(\vec{k}\,^2+\kappa^2)^2}\ , 
\end{equation} 
where $\kappa=\sqrt{m|\epsilon_b|}=m\alpha/2$. It is normalized,  however,
according to (\ref{nor5}) with $\varepsilon_k\approx m$ and $N_2=1$. The 
integral over $d^3k'$ in (\ref{eqwf3}) is concentrated in the region 
$k'\approx \kappa$.  Therefore, at $k\gg \kappa$ the momentum  $\vec{k}\,'$ in
$V(\vec{k}\,',\vec{k},\vec{n},M^2)$ can be ignored,  and  from (\ref{eqwf3}) we
find:   
\begin{equation}\label{k7} 
\psi(\vec{k},\vec{n})=-\frac{mV(0,\vec{k},\vec{n},M^2)}                
{(2\pi)^3(\vec{k}\,^2
+\kappa^2)}\int\psi(\vec{k}\,')d^3k'\ .                     
\end{equation}                                                                  
Substituting in the r.h.s. of eq.(\ref{k7}) the expressions (\ref{k3},\ref{k4})
for $V$ and (\ref{k6}) for $\psi$, we obtain                
\begin{equation}\label{k8}                                                     
\psi(\vec{k},\vec{n})                                                           
=\frac{8\sqrt{\pi m}\kappa^{5/2}}{(\vec{k}\,^2+\kappa^2)^2 
\left(1+\frac{\displaystyle |\vec{n}\cd\vec{k}|}                                   
{\displaystyle\varepsilon_k}\right)}\ .                                         
\end{equation}                                                                  
This relativistic wave function of the ground state with zero  total angular
momentum is a good approximation of a more exact one in the range  $k>\kappa$.
Corrections of order $\alpha \log(\alpha)$ should be considered in  the range
$k<\kappa$ (see \cite{fft73}). Though the kernel (\ref{k3}), (\ref{k4}) 
contains  the modulus $|\vec{n}\cd\vec{k}\,'/\varepsilon_{k'} 
-\vec{n}\cd\vec{k}/\varepsilon_k|$, one can show that the exact solution of
(\ref{eqwf3}) has no ``cusp'' at $\vec{n}\cd\vec{k}=0$. This cusp in
(\ref{k8}) appears due to our approximations.  

One can check in this simple example that it is the retardation of  the 
interaction that is the dynamical reason for the dependence of the  wave 
function on the variable $\vec{n}$. The non-relativistic Coulomb  expression
for the kernel (\ref{k5}) does not contain retardation and  does not depend on
$\vec{n}$ while the relativistic kernel (\ref{k3})  contains retardation and
depends on $\vec{n}$. This leads to the  dependence of the wave function on the
argument $\vec{n}$. 

The retardation leads to both the $\vec{n}$-dependence  and the presence  of
the carriers of the  interaction in the intermediate state, which contribute to
the many body sectors.  However,  these  two effects, being important in full
measure in a truly relativistic  system, can manifest themselves in a different
way in weakly bound  systems. Neglecting the many-body sectors does not
necessarily entails  to neglect the $\vec{n}$-dependence of the  wave function
at $k \approx m$.  It is  necessary to take into account   the
$\vec{n}$-dependence of the wave function even when one  restricts to the
two-nucleon sector. 

We emphasize that the dependence of the wave function (\ref{k8}) on $\vec{n}$
does not mean any violation of the rotational invariance. As explained above, 
it reflects the   dependence (unavoidable one, in the field-theoretical
framework) of any off-energy shell amplitude on the orientaion of the
light-front plane. At the same time, the on-shell amplitude expressed through
the wave function should not depend on $\vec{n}$.  For the case of
electromagnetic form factor this property is discussed below  in sect.
\ref{emff}. 

The wave  function of the 2p state can be found analogously. In the 
system where $\vec{k}_1+\vec{k}_2=0$ it has the form \cite{karm80}:                                                                  
\begin{eqnarray}\label{bs12}                                                   
&&\psi ^\lambda (\vec k,\vec n)= \frac{8\pi \kappa 
^{7/2}m^{1/2}}{\sqrt{6}}\frac {1}{\left(\vec k\,^2+\frac{1} 
{\displaystyle 4}\kappa ^2\right)^3\left(1+ \frac{\displaystyle |\vec 
n\cd\vec k|}{\displaystyle\varepsilon_k}\right)^2} 
\\ 
&&\times
\left\{kY_{1\lambda }(\vec k/k) +Y_{1\lambda}(\vec{n}) 
\left[\frac{(2\varepsilon_k-M)^2} {4\varepsilon_k M} 
(\vec n\cd\vec k)-\frac{(\vec k^2+\frac{\displaystyle 1}
{\displaystyle 4}\kappa ^2)} {2m}\left(\theta (-\vec n\cd\vec 
k)-\theta (\vec n\cd\vec 
k)\right)\right]\right\}.
\nonumber
\end{eqnarray} 
The wave function corresponding to the angular  momentum  $l=1$ contains the
spherical function $Y_{1\lambda }(n)$. This is  an illustration of the fact 
that the vector $\vec n$ participates in  the construction of the total angular
momentum on the same ground  as the relative momentum $\vec{k}$.   The
dynamical difference between the  solution with  $\vec{k} \| \vec{n}$ and
$\vec{k} \perp \vec{n}$ is obviously related  to the  property that some of the
components of the angular momentum  $\vec{J}$, before using the angular
condition, depend on the interaction.

%%%%%%%%%%%%%%%%%%%%%%%%%%%%%%%%%%%%%%%%%%%%%%%%%%%%%%%%%%%%%%%%%%%%%
         
\subsection{Relation with the Bethe-Salpeter function}\label{bsf}                       
It is instructive to compare the solution (\ref{k8}) with one found
using the Bethe-Salpeter function. 

Fisrt, we  find the relation between the light-front wave function and the 
Bethe-Salpeter function. We should start from the integral that  restricts the
variation of the arguments of the Bethe-Salpeter  function to the light-front
plane:                                                     
\begin{equation} \label{bs1}I=\int                                              
d^4x_1\ d^4x_2\ \delta (\omega \cd x_1)\ \delta (\omega\cd x_2)\                     
\Phi(x_1,x_2,p)\exp (ik_1\cd x_1+ik_2\cd x_2)\ ,                                      
\end{equation} where $k_1,k_2$ are the on-shell momenta:
$k_1^2=k_2^2=m^2$,  and $\Phi (x_1,x_2,p)$ is the                                  
Bethe-Salpeter function \cite{bs}, eq.(\ref{bs2}).                                                                   
We represent the $\delta $-functions in (\ref{bs1}) by the integral form
$$\delta(\omega\cd x)=
\frac{1}{2\pi}\int \exp(-i\omega\cd x\alpha)d\alpha,$$
introduce the Fourier transform of the Bethe-Salpeter function $\Phi (k,p)$, 
$$ \Phi(x_1,x_2,p)=(2\pi )^{-3/2}\exp \left[-ip\cd 
(x_1+x_2)/2\right]\tilde{\Phi          
}(x,p)\ ,\quad x=x_1-x_2\ ,$$                                                   
$$                                                               
\Phi (l,p)=\int \tilde \Phi (x,p)\exp (il\cd x)d^4x\ ,                   
$$                                                                 
where $l=(l_1-l_2)/2$, $p=l_1+l_2$, $l_1$ and $l_2$ are off-mass  shell 
four-vectors, and make the change of variables  $\alpha_1+\alpha_2=\tau$,
$(\alpha_2-\alpha_1)/2=\beta$.                                                                  

On the other hand, the integral (\ref{bs1}) can be expressed in terms  of the
two-body light-front wave function. We assume that the  light-front plane is
the limit of a space-like plane, therefore  the  operators $\varphi (x_1)$ and
$\varphi (x_2)$ commute, and, hence,  the  symbol of the $T$ product in
(\ref{bs2}) can be omitted. In the  considered representation, the Heisenberg
operators $\varphi (x)$ in  (\ref{bs2}) are identical on the light front
$\omega\cd x=0$ to the  Schr\"odinger operators (just as in the ordinary
formulation of field  theory the Heisenberg and Schr\"odinger operators are
identical for  $t=0$).  The Schr\"odinger operator $\varphi (x)$ (for the
spinless  case for simplicity), which for $\omega\cd x=0$ is the free field 
operator, is given by (\ref{ft1}). We represent the state vector $|p\rangle
\equiv \phi (p)$ in  (\ref{bs2}) in the form of the expansion (\ref{wfp1}).
Since the vacuum  state on the light front is always ``bare'', the creation
operator,  applied to the vacuum state $\langle 0|$ gives zero, and in the 
operators $\varphi (x)$  the part containing the annihilation  operators only
survives. This cuts out the two-body Fock component in  the  state vector. We
thus obtain:  
\begin{equation} \label{bs6} 
I=\frac{(2\pi )^{3/2}(\omega\cd p)}{2(\omega\cd k_1)(\omega\cd 
k_2) }\int_{-\infty }^{+\infty }\psi (k_1,k_2,p,\omega \tau )\delta 
^{(4)}(k_1+k_2-p-\omega \tau )\ d\tau\ .  
\end{equation} 
Comparing (\ref{bs1}) and (\ref{bs6}), we find:
\begin{equation} \label{bs8}                                                     
\psi(k_1,k_2,p,\omega \tau ) =\frac{(\omega\cd k_1 )(\omega\cd k_2 
)}{\pi (\omega\cd  p)}\int_{-\infty }^{+\infty }\Phi 
(l_1=k_1-\omega\tau/2+\omega\beta,l_2 
=k_2-\omega\tau/2-\omega\beta,p)d\beta       
\end{equation}      
where $\Phi(l_1,l_2)$ is the Bethe-Salpeter function  parametrized in terms of
the off-mass shell momenta $l_1,l_2$.   The argument $p$ in(\ref{bs8}) is
related to the on-shell  momenta $k_1,k_2$ as $p=k_1+k_2-\omega\tau$, in
contrast to off-mass shell relation $p=l_1+l_2$.

In ordinary LFD, eq.(\ref{bs8})  corresponds to the integration over $dk_{-}$.
This equation makes the  link between the Bethe-Salpeter function $\Phi$ and
the wave function  $\psi$ defined on the light front specified by $\omega$.  It
should be  noticed however that eq.(\ref{bs8}) is not necessarily an exact
solution of  eq.(\ref{eqwf3}), since, as a rule, different approximations are
made for the Bethe-Salpeter kernel and for the light-front one. In the ladder
approximation, for example, the Bethe-Salpeter amplitude contains the box 
diagram, including the time-ordered diagram with two exchanged particles in the
intermediate state, as indicated in graphically in eq. (\ref{sup1}) in sect
\ref{hfst}. This contribution is absent in the  light-front ladder kernel.

Note also the interesting paper \cite{mitra98}, (for earlier studies see
\cite{mitra92}), where the Markov-Yukawa transversality principle for the
two-body Bether-Salpeter kernel was formulated on the covariant light-front
plane.  It allows not only to obtain an exact three-dimensional reduction of
the Bethe-Salpeter equation,  but also to make the exact reconstruction of the
four-dimensional  Bethe-Salpeter equation from the three-dimensional form. The
three-dimensional form is convenient for spectroscopical calculations, the
four-dimensional form facilitates the evaluation of the loop integrals for the
form factors.  In particular cases  the methood gives the same results as
obtained earlier by other description  \cite{mitra87,chak89}. A three-quark
generalization is given in \cite{mitra3q}.

The quasipotential type equations for the light-front wave function derived by
restricting arguments of the Bethe-Salpeter amplitude to the light-front plane
$z+t=0$ and corresponding electromagnetic form factors were studied  in refs.
\cite{gkmtf75,gm75}.

%%%%%%%%%%%%%%%%%%%%%%%%%%%%%%%%%%%%%%%%%%%%%%%%%%%%%%%%%%%%%%%%%%%%%
%%         
\subsection{Solution in the Bethe-Salpeter approach}\label{wcbs}                                                                    
The exact expression for the Bethe-Salpeter function in the Wick-Cutkosky model
is found in the form of the integral  representation \cite{wcm,nak69} and,
for zero angular momentum, reads:                       
\begin{equation}                                                                
\label{bs8p}                                                                    
\Phi (l,p)=-\frac i{\sqrt{4\pi }}\int_{-1}^{+1}\frac{               
g(z,M)dz}{(m^2-M^2/4-l^2-zp\cd l-i\epsilon)^3}\ .                                   
\end{equation}                                                                  
The spectral function $g(z,M)$ is determined by a differential equation \cite
{wcm,nak69} and has no singularity at $z=0$. The approximate explicit
solution found in \cite{wcm} for $g(x,M)$ has the  form:  
\begin{equation}\label{bs11} 
g(z,M)=2^6\pi \sqrt{m}\kappa ^{5/2}(1-|z|)\ .  
\end{equation} 
The discontinuity of the spectral function $g(z,M)$ at $z=0$ is a result of
approximation, since the solution (\ref{bs11})  corresponds to an
asymptotically small binding energy. Inserting (\ref{bs11}) in (\ref{bs8p}) and
integrating over $z$,   one can recover the solution of the Bethe-Salpeter 
equation:
\begin{equation}\label{wcm02}
\Phi(k,p)=-ic\left[\left(m^2-\frac{1}{2}M^2-k^2\right)
\left(m^2-(\frac{1}{2}p+k)^2-i0\right)
\left(m^2-(\frac{1}{2}p-k)^2-i0\right)\right]^{-1},
\end{equation}
where $c=2^5\sqrt{\pi m\kappa^5}$ with $\kappa=\sqrt{m|\epsilon_b|}= m\alpha
/2$. 

To find the light-front wave function, one can substitute in eq.(\ref{bs8})
the Bethe-Salpeter function either in the form (\ref{bs8p}) or in (\ref{wcm02}).
From (\ref{bs8p}) we find  \cite{bjs}:         
\begin{equation}\label{bs10}
\psi =\frac{g(1-2x,M)}{2^5\sqrt{\pi }x(1-x) 
(\vec k\,^2+\kappa ^2)^2}\ .                                                                       
\end{equation}  
                                                                
Substituting (\ref{bs11}) in (\ref{bs10}), we reproduce the expression
(\ref{k8})  for the light-front wave function. 

%%%%%%%%%%%%%%%%%%%%%%%%%%%%%%%%%%%%%%%%%%%%%%%%%%%%%%%%%%%%%%%%%%%%%

\subsection{Including spin}\label{spin} 

As explained in sect. \ref{frform}, in the standard version of LFD  the
generators of the Poincar\'e group coresponding to the Lorentz boosts changing
the orientation of the plane $t+z=0$, are the dynamical ones and contain the 
interaction. In the explicitly covariant version of LFD the dependence of the
wave function on the light-front  orientation is taken into account by means of
the variable $\omega$. Now, using kinematics (i.e., the transformation
properties) we have to ensure that this wave function corresponds to a definite
total angular momentum. In the case of the zero angular momentum the
four-vector $\omega$ enters alway in the scalar product with the particle
four-momenta. For the non-zero spins $\omega$ appears in the spin structures. 

We illustrate the construction of the states with spins by two examples.

Consider a system consisting of  quark and antiquark  in the $J^{\pi}=0^-$
state ("pion"). The light-front wave function  has the form:  
\begin{equation}\label{pion}                                                    
\psi=\bar{u}(k_2)\left[A_1\frac{1}{m}+ 
A_2\frac{\hat{\omega}}{\omega\cd p}\right]
\gamma_5 v(k_1) ,
\end{equation}
where $\bar{u}$ and $v$ are the spinors, 
$\hat{\omega}=\omega_{\mu}\gamma^{\mu}$, $A_{1,2}$ are the scalar functions,
$m$ is the quark mass.  In the system of reference  where
$\vec{k}_1+\vec{k}_2=0$ this wave function obtains the form:
\begin{equation}\label{eqp3}
\psi=w_2^t\left(g_1+\frac{i\vec{\sigma}\cd [\vec{n}\times 
\vec{k}]}{k}g_2\right)w_1\ ,
\end{equation} 
with the following relations between the invariant functions:
$$
A_1=-\frac{m}{2 \varepsilon_k} (g_1+\frac{m}{k} g_2), \quad 
A_2=\frac{\varepsilon_k}{k} g_2\ .
$$
Note that there exists a special representation (see \cite{cdkm}) in which the
wave function has the form (\ref{eqp3}) in arbitrary system of reference.

From eqs.(\ref{pion},\ref{eqp3}) one can see that the spin structure of the
wave function indeed contains the four-vector $\omega$ determining the
light-front orientation. Due to that  it is determined by two invariant
functions. Only one of them ($g_1$) survives in the nonrelativistic limit.

Another example is the light-front wave function of a system consisting of two
fermions in the state with total angular momentum equal to 1. This can be two
nucleons in the state $J^{\pi}=1^+$ (the deuteron) or the  quark-antiquark pair
in the state $J^{\pi}=1^-$ ($\rho$-meson). This wave function has the form:
\begin{equation}\label{nz1}                                                     
{\mit\Phi}^{\lambda}_{\sigma_2\sigma_1}(k_1,k_2,p,\omega \tau)=
\sqrt{m}e^{\lambda}_{\mu}(p) 
\bar{u}^{\sigma_2}(k_2)\phi_{\mu}U_c\bar{u}^{\sigma_1}(k_1)\ ,        
\end{equation}                                                                  
with                                                                            
\begin{eqnarray}\label{nz2} \phi_{\mu}&=&
\varphi_1\frac{(k_1- k_2)_{\mu}}{2m^2}
+\varphi_2\frac{1}{m}\gamma_{\mu}                           
+\varphi_3\frac{\omega_{\mu}}{\omega\cd p}                                       
+\varphi_4\frac{(k_1-k_2)_{\mu}\hat{\omega}}{2m\omega\cd p}
\nonumber\\ 
&-&\varphi_5\frac{i}{m^2\omega\cd p}\gamma_5
\epsilon_{\mu\nu\rho\gamma}          
k_{1\nu}k_{2\rho}\omega_{\gamma}
+\varphi_6\frac{m\omega_{\mu}\hat{\omega}}{(\omega\cd p)^2} .                      
\end{eqnarray}                                                                                                                          
It is determined by six invariant functions $\varphi_{1-6}$, depending on two
scalar variables. This number  is the dimension of the matrix depending on the
spin projections of the deuteron and two nucleons, divided by the factor 2 due
to the parity  conservation: $N=3\times 2 \times 2/2=6$.

In the system of reference where $\vec{k}_1+\vec{k}_2=0$ (or in arbitrary
system, but in the representation described in \cite{cdkm})  this wave
function obtains the form:
\begin{equation}\label{nz7}                                                     
{\mit\Psi}^{\lambda}_{\sigma_2\sigma_1}(\vec{k},\vec{n}) = 
\sqrt{m}w^\dagger_{\sigma_2} \psi^{\lambda}(\vec{k},\vec{n})\sigma_y 
w^\dagger_{\sigma_1}\ , 
\end{equation} 
with 
\begin{eqnarray}\label{nz8} 
\vec{\psi}(\vec{k},\vec{n}) & = & f_1\frac{1}{\sqrt{2}}\vec{\sigma} + 
f_2\frac{1}{2}(\frac{3\vec{k}(\vec{k}\cd\vec{\sigma})}{\vec{k}^2} 
-\vec{\sigma}) + f_3\frac{1}{2}(3\vec{n}(\vec{n}\cd\vec{\sigma}) 
-\vec{\sigma}) \nonumber \\ & + & 
f_4\frac{1}{2k}(3\vec{k}(\vec{n}\cd\vec{\sigma}) + 
3\vec{n}(\vec{k}\cd\vec{\sigma}) - 2(\vec{k}\cd\vec{n})\vec{\sigma}) 
\nonumber        
\\ & + & f_5\sqrt{\frac{3}{2}}\frac{i}{k}[\vec{k}\times \vec{n}] +              
f_6\frac{\sqrt{3}}{2k}[[\vec{k}\times \vec{n}]\times\vec{\sigma}]\ ,            
\end{eqnarray} 
where $w$ is the two-component nucleon spinor normalized to $w^\dagger w=1$.
The relations between $\varphi$ and $f$ can be found in \cite{cdkm}. In the
relativistic  one boson exchange model this wave function was calculated in
\cite{ck-deut}.  It was found that the function $f_5$, of relativistic origin,
is very important: it dominates at $k > 500$ MeV/c.  In nonrelativistic the
functions $f_{3-6}$ become negligible, and  only two first structures survive,
corresponding to usual S- and D-waves.

This wave function was used in the paper \cite{ck99} to calculate the deuteron
electromagnetic form factors. No any parameters were fitted.  It turned out
that the calculated structure function $A(Q^2)$ and  the polarization
observable $t_{20}$ coincide with rather precise experimental data obtained
recently at CEBAF/TJNAF.

%%%%%%%%%%%%%%%%%%%%%%%%%%%%%%%%%%%%%%%%%%%%%%%%%%%%%%%%%%%%%%%
\subsection{The nucleon wave function}\label{nwf}

Many calculations of the nucleon properties (magnetic moments, form  factors,
etc.)  are carried out in the framework of  LFD, in the three-quark model, with
the nucleon wave function containing one or a few spin components.  The total
number of the spin components in the nucleon wave function is sixteen
\cite{karm98}. This  is related to the fact known long ago \cite{kolyb65} that
in a  many-body system the parity conservation does not reduce the number of 
the spin components. This is so for a relativistic three-body system  and for
any $n$-body system for $n\geq 4$ (both relativistic and  nonrelativistic one).
Hence, for the relativistic nucleon we get
$$
N=(2S_1+1)(2S_2+1)(2S_3+1)(2S_N+1)
 =2\times 2\times 2\times 2=16.
$$ 
These 16 components are forming the full basis for the nucleon wave 
function.

In  nonrelativistic limit the parity conservation reduces this  number down to
8 components. Their relativistic counterparts were found is \cite{dz88}. Note,
however, that one can construct also another 8 components with the opposite
parity.

The difference between relativistic and nonrelativistic cases is related to the
fact that in relativistic case one can construct  the  pseudoscalar:
\begin{equation}\label{eq2}
C_{ps}=\epsilon^{\mu\nu\rho\gamma}k_{1\mu}k_{2\nu}k_{3\rho}p_{\gamma}.
\end{equation}
It is not zero, since the bound quarks are off-energy-shell:  $k_1
+k_2+k_3=p+\omega \tau\neq p.$ In ordinary light-front approach this
corresponds to  the well known conservation law:
$$\vec{k}_{1\perp} 
+\vec{k}_{2\perp}+\vec{k}_{3\perp}=\vec{p}_{\perp},\quad
k_{1+} +k_{2+}+k_{3+}=p_{+},\quad
\mbox{but}\quad k_{1-} +k_{2-}+k_{3-}\neq p_{-}.$$
Therefore, we can take 8 componets with opposite parity, multiply them by
$C_{ps}$ and get another 8 componets with the nucleon parity. By this way, we
get 16 components of the nucleon wave function.  They are given in
\cite{karm98}. Due to the momentum conservation, the pseudoscalar (\ref{eq2})
can be  rewritten as:
$$
C_{ps}=-\tau\epsilon^{\mu\nu\rho\gamma}
k_{1\mu}k_{2\nu}k_{3\rho}\omega_{\gamma}.
$$
It is proportional to $\omega$. So, namely the dependence of the relativistic
nucleon wave function on the light-front orientation $\omega$ is the reason of 
appearence of 8 extra componets. In nonrelativistic case this dependence 
disappears, and we remain with 8 components. Formally, this is due to the fact,
that $\omega$ enters in the momentum conservation law in the  combination
$\omega\tau$, where   $\tau =(s-M^2)/(2\omega\cd p)$. This term contains extra
factor $k/m$ and disappears at $k\ll m$. We get the nonrelativistic
conservation law: $\vec{k}_1+\vec{k}_2+\vec{k}_3=\vec{p}$ and loose opportunity
to construct any pseudoscalar and the extra components.

As mentioned above, an advantage of the explicitly covariant LFD is
simplification of the  transformation properties of the wave functions with a
given spin. In the standard LFD approach the wave function is transformed in
every spin index by a special  Melosh rotation matrices \cite{melosh}. In the
covariant version, the transformation properties are automatically taken into
account and do not require any Melosh matrices.

Consider, for example, the nucleon wave function in c.m.-system with fully
symmetric S-wave spin-isospin structure (implicitly multiplied by the
antisymmetric color singlet function):
\begin{equation}\label{eqwf}
\Psi_S= \frac{\psi_S}{\sqrt{72}}
[3+(\vec{\sigma}_{12}\cdot\vec{\sigma}_{3N})
(\vec{\tau}_{12}\cdot\vec{\tau}_{3N})],
\end{equation}
where $\vec{\sigma}_{12}=(w_1^{\dagger}\vec{\sigma}\sigma_{y}w_{2}),  \quad
\vec{\sigma}_{3N}=(w_3^{\dagger}\vec{\sigma}\sigma_{y}w_{N})$  and similarly
for the isospin matrices $\vec{\tau}_{12}$,  $\vec{\tau}_{3N}$. Using the Fierz
identities, one can check that the wave function (\ref{eqwf}) is indeed
symmetric relative to permutation of any  pair of quarks (provived $\psi_S$ is
symmetric). In arbitrary system it is multiplied by the Melosh rotation
matrices. For $\psi_S$ one can take, for example, the harmonic oscillator
model:
$$
\psi_S=\frac{2^4{\pi}^{3/2}3^{1/4}N}{\alpha^3}
\exp\left(-\frac{\vec{k}_1\,^2+\vec{k}_2\,^2+\vec{k}_3\,^2}
{2\alpha^2}\right),
$$
$\vec{k}_i$ are the quark relative momenta,  $N$ is a normalisation factor
equal to 1 in the nonrelativistic limit.

In the explicitly covariant LFD it is represented it in covariant, 
four-dimensional  form, in terms of the usual Dirac spinors,  avoiding any
Melosh matrices. For this aim we introduce the projection operators:
$$
\Pi_+=\frac{{\cal M}+\hat{\cal P}}{2{\cal M}},\quad
\Pi_-=\frac{{\cal M}-\hat{\cal P}}{2{\cal M}},
$$
where $U_c= \gamma^2\gamma^0$ is the charge conjugation matrix, ${\cal
P}=k_1+k_2+k_3=p+\omega\tau, \quad  \hat{\cal P}=\gamma^{\mu}{\cal
P}_{\mu},\quad {\cal M}^{2}={\cal  P}^{2}.$  ${\cal M}$ here is the effective
mass of  the free quarks (not the nucleon mass).  Then the wave function
(\ref{eqwf}) is covariantly represented  as \cite{karm98}:
\begin{eqnarray}\label{eq86p} 
{\mit\Psi}_S&=&\frac{\psi_S}{\sqrt{72}}c_1 c_2 c_3 c_N 
\{3[\bar{u}(k_1)\Pi_+\gamma_5  U_c\bar{u}(k_2)] 
[\bar{u}(k_3)\Pi_+ u(p)]  
\nonumber\\
&-&[\bar{u}(k_1)\Pi_+\gamma^{\mu}\Pi_- U_c\bar{u}(k_2)]
[\bar{u}(k_3)\Pi_+\gamma_{\mu}\gamma_5 \Pi_+ u(p)]
(\vec{\tau}_{12}\cdot\vec{\tau}_{3N})\},
\end{eqnarray}
where $c_{1,2,3}=1/\sqrt{\epsilon_{1,2,3}+m},\quad 
c_{N}=1/\sqrt{\epsilon_{N}+M}$ and, e.g., 
$\epsilon_{1}=\sqrt{\vec{k}_{1}^{2}+m^{2}}$ is the energy of the quark  1.  In
the system where $\vec{k}_1+\vec{k}_2+\vec{k}_3=0$ this wave  function {\it
exactly} coincides with (\ref{eqwf}).  The wave function (\ref{eq86p}) can be 
decomposed in terms of the 16 structures discussed above. Other  states are
represented similarly. The calculation of the nucleon properties  (magnetic
moments, electromagnetic form factors, etc.) is now a standard routine using
the trace  techniques of the Dirac matrices.  In comparison to the standard
light-front approach, for the identical nucleon wave functions, the resuts in
both approaches coincide with each other, but in the explicitly covariant
approach they are obtained much more simpler.

%%%%%%%%%%%%%%%%%%%%%%%%%%%%%%%%%%%%%%%%%%%%%%%%%%%%%%%%%%%%%%%%%%%%%
          
\section{Electromagnetic form factors}\label{emff}
The general physical electromagnetic amplitude of a spinless system is  given
by: 
\begin{equation}\label{emv2}                                                
J_{\rho}\equiv\langle p'|J_{\rho}(0)|p\rangle =(p+p')_{\rho}F(Q^2)\ .                         
\end{equation}  
where $F(Q^2)$ is the electromagnetic  form factor.
In LFD it is obtained by calculating the amplitude corresponding to fig.
\ref{f1-ks92-b}:
               
\begin{figure}[htbp]
\centerline{\epsfbox{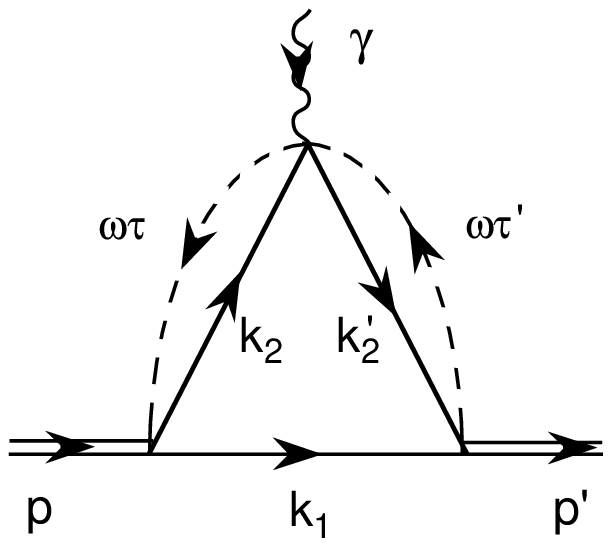}}
\figcap{Electromagnetic vertex of a bound system.}
\label{f1-ks92-b}
\end{figure}  

\begin{equation}\label{emv4}                                                    
J_{\rho}=\frac{1}{(2\pi)^3}\int\frac{(p+p'+\omega\tau +\omega\tau' 
-2k_1)_{\rho}}{(1-\omega\cd k_1/\omega\cd p)^2}\ \psi' \psi\ 
\theta\left(\omega\cd (p-k_1)\right) \frac{d^3k_1}{2\varepsilon_{k_1}}\ .          
\end{equation}
 
Exact light-front amplitude on the energy shell has to coincide with the
Feynman one and should not depend on the orientation of the light front plane.
It should reproduce the form (\ref{emv2}). However, the diagram \ref{f1-ks92-b}
corresponds to impulse approximation,  when the electromagnetic current does
not contain any interaction. Therefore the dependence of amplitude (\ref{emv4})
on the light-front orientation survives. $J_{\rho}$ depends  on $\omega$. It
also can be represented in the general form: 
\begin{equation}\label{emv1}                                                    
J_{\rho}=(p+p')_{\rho}F(Q^2) + \frac{\omega_{\rho}}{\omega\cd p} B_1(Q^2)\ .  
\end{equation} 
The factor $1/\omega\cd p$ is separated for convenience. The invariant 
functions $F$ and $B_1$  depend on $Q^2=-q^2 \equiv -(p'-p)^2$. They could
depend in  principle on $\omega \cd p$ and $\omega\cd p'$.  However, the 
four-vector $\omega$ is defined up to an arbitrary number, and, hence,  the
theory is invariant relatively to the replacement $\omega\rightarrow 
\alpha\omega$, where $\alpha$ is a number. The form factors $F$ and  $B_1$ can
therefore depend only on the ratio $\omega \cd p'/\omega\cd  p$.

Now we take into account that $\omega$ is restricted by the condition
$\omega\cd q=0$, implying the transversality of $q$. In this case we have
$\omega\cd p'/\omega\cd p =1$, and the functions $F$ and $B_1$  depend on $Q^2$
only.                 

The main difference of the amplitude (\ref{emv1}) with respect to  (\ref{emv2})
is the presence of an additional contribution, proportional to 
$\omega_{\rho}$. To avoid any misunderstanding, we emphasize  that even {\em in
the case  where the wave function $\psi$ does not depend on $\vec{n}$}, the
term  proportional to $\omega_{\rho}$  still survives in the electromagnetic
vertex.

In the spinless case, the  physical  form factor, $F(Q^2)$ can be obtained
immediately by multiplying both sides of eq.(\ref{emv1}) by $\omega_{\rho}$. We
thus get:                       
\begin{equation}\label{emv9} 
F(Q^2)=\frac{J \cd\omega}{2\omega\cd p}\ .          
\end{equation}                                                                  
With (\ref{emv4}), (\ref{emv9}) we obtain:
\begin{equation}\label{ff0}
F(Q^2)= \frac{1}{(2\pi)^3} \int \psi(\vec{R}_{\perp}^2,x)
\psi((\vec{R}_{\perp}-x\vec{\Delta})^2,x)\frac{d^2R_{\perp}dx}{2x(1-x)}.
\end{equation}
We have represented here, and in the following, the four-momentum transfer $q$ 
by $q=(q_0,\vec{\Delta},\vec{q}_{\|})$ with  $\vec{\Delta}\cd\, \vec{\omega}=0$
and $\vec{q}_{\|}$ is parallel to  $\vec{\omega}$. Since $\omega \cd q=0$, we
have $Q^2=-q^2=\vec{\Delta}^2$.

The form factor in the Bethe-Salpeter approach is found from the formula:
\begin{equation}\label{wcm01}
(p+p')_{\rho}F(Q^2)=i\int (p+p'-2k)_{\rho}\Phi(\frac{1}{2}p-k,p)
\Phi(\frac{1}{2}p'-k,p')(m^2-k^2)\frac{d^4k}{(2\pi)^4}.
\end{equation}
In the Wick-Cutkosky model, for instance, the light-front wave function is
given by eq.(\ref{k8}) and  the Bethe-Salpeter function $\Phi(k,p)$ is given by
eq. (\ref{wcm02}). {\em The form factors calculated by means of both approach
coincide with  each other with high accuracy}. Both approaches give the same
asymptotical behavior of the form factors at $|t|=Q^2\gg m^2$:
$$
F(t)\approx \frac{16\alpha^4m^4}{t^2}\left[1+
\frac{\alpha}{2\pi}\log\left(\frac{|t|}{m^2}\right)\right],
$$
where $\alpha=g^2/(16\pi m^2)$, $g$ is the coupling constant in the 
Wick-Cutkosky model. 

In the usual light-front formulation, with $\omega =(1,0,0,-1)$,
eq.(\ref{emv9}) corresponds to expressing the form factor through  the $J_+$
component. This is well known, and  eq.(\ref{ff0}) has been found in ref.
\cite{gbb72}. However,  this procedure cannot be extended to the calculation of
physical  form factors of systems with total spin 1/2 and 1. Their
electromagnetc vertices also depend on the four-vector $\omega$.

For for a spin-1 particle this vertex has the form:
\begin{equation}\label{ff4} 
\langle\lambda'|J_\rho|\lambda\rangle =
\frac{\displaystyle{1}}{\displaystyle{2\omega\cd p}}
e_{\mu}^{*\lambda'}(p')J^{\mu\nu}_{\rho}e_{\nu}^{\lambda}(p),
\quad \mbox{where} \quad
J^{\mu\nu}_{\rho}= T^{\mu\nu}_{\rho} + 
B^{\mu\nu}_{\rho}(\omega) .
\end{equation}
Here $T^{\mu\nu}_{\rho}$ is  determined by the physical form factors and has
the usual  structure \cite{gj57}:         
\begin{eqnarray}\label{ff1}                                                     
\langle\lambda'|J_\rho|\lambda\rangle &=& 
e_{\mu}^{*\lambda'}(p')\left\{P_{\rho}\left[{\cal F}_1(q^2) 
g^{\mu\nu} + {\cal F}_2(q^2){q^{\mu}q^{\nu}\over 2M^2}\right]
\right.                                                                                                 
+ \left.{\cal G}_1(q^2) (g^{\mu}_{\rho}q^\nu - 
g^{\nu}_{\rho}q^\mu)\right\}e_{\nu}^{\lambda}(p)\nonumber\\ 
&\equiv& 
e_{\mu}^{*\lambda'}(p')T^{\mu\nu}_{\rho}e_{\nu}^{\lambda}(p),              
\end{eqnarray}                                                                  
$e_{\mu}^{\lambda}(p)$ is the spin-1 polarization vector, $p$ 
and $p'$ are the initial and final momenta,
$\lambda$ and $\lambda'$ are the corresponding helicities, 
$P=p+p'$ and $q = p' - p$.
 The tensor
$B^{\mu\nu}_{\rho}$ contains the $\omega$  dependent 
terms:                                                                          
%%%%%%%%%%%%%%%%%%%%%%%%%%%                                                    
\begin{eqnarray}\label{ff5}                                                     
B^{\mu\nu}_{\rho} &=& {M^2\over2(\omega\cd
p)}\;\omega_{\rho}\left[B_1 g^{\mu\nu} +
B_2{q^{\mu}q^{\nu}\over M^2} +  B_3 M^2
{\omega^{\mu}\omega^{\nu}\over (\omega \cd p)^2}
+  B_4 {q^{\mu}\omega^{\nu} -
q^{\nu}\omega^{\mu}\over2\omega\cd p}  \right]       
\nonumber\\                                                                     
&+& B_5 P_{\rho} M^2 {\omega^{\mu}\omega^{\nu}\over (\omega \cd p)^2 } 
+ B_6 P_{\rho} {q^{\mu}\omega^{\nu} - q^{\nu}\omega^{\mu} \over 
2\omega\cd p} + B_7 M^2 {g^{\mu}_{\rho}\omega^{\nu} + 
g^{\nu}_{\rho}\omega^{\mu} \over \omega \cd p} 
\nonumber\\                                                                
&+& B_8 q_{\rho} {q^{\mu}\omega^{\nu} + q^{\nu}\omega^{\mu}\over 
2\omega \cd p},
\end{eqnarray}                                                          
%%%%%%%%%%%%%%%%%%%%%%%%%%%                                                     
$B_1,...,B_8$  are invariant functions.  This tensor  is not eliminated by
contraction with  $\omega_{\rho}$.  In these cases the electromagnetic form
factors are given by contraction of the electromagnetic vertex with more
complicated tensors found in \cite{ks92,ks94}. The current component $J_+$ is
still enough to find the form factors  ${\cal F}_1,{\cal F}_2$, but it is not
enough to find ${\cal G}_1$.

The formulas for the physical form factors for the case of spin-1/2 
light-front electromagnetic vertex (nucleon electromagnetic form factors, for
instance) are found in \cite{km96}.

%%%%%%%%%%%%%%%%%%%%%%%%%%%%%%%%%%%%%%%%%%%%%%%%%%%%%%%%%
\section{Suppression of the higher Fock states}\label{hfst}
The kernel corresponding to exchange by a particle in the Bethe-Salpeter
approach and in LFD are not equivalent to each other. The light-front graphs
are obtained from the Feynman ones by time-ordering of the vertices. For
example, the Feynman  graph with two exchanges corresponds to the following sum
of the time ordered graphs:
\begin{eqnarray}
\raisebox{-.3cm}{\epsfxsize=1.4cm \epsffile{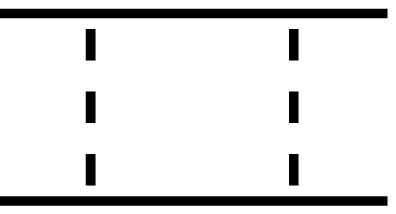}}
=
\raisebox{-.25cm}{\epsfxsize=1.4cm \epsffile{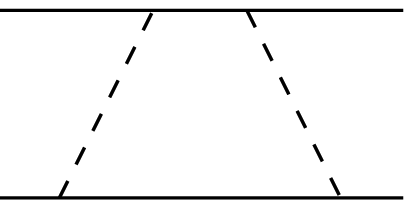}}
+
\raisebox{-.25cm}{\epsfxsize=1.4cm \epsffile{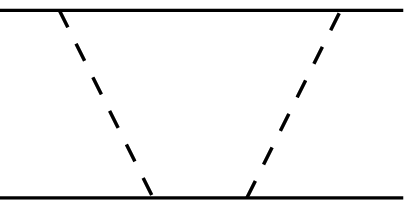}}
+
\raisebox{-.25cm}{\epsfxsize=1.4cm \epsffile{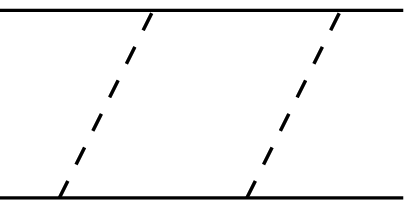}}\nonumber\\
+
\raisebox{-.25cm}{\epsfxsize=1.4cm \epsffile{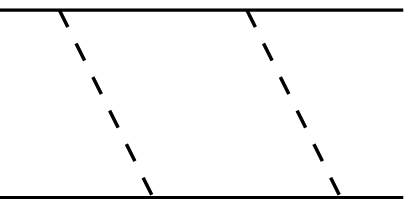}}
+
\raisebox{-.25cm}{\epsfxsize=1.4cm \epsffile{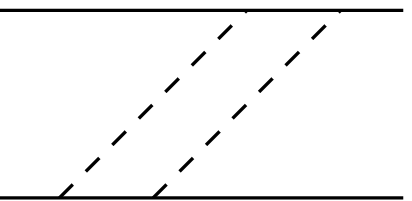}}
+
\raisebox{-.25cm}{\epsfxsize=1.4cm \epsffile{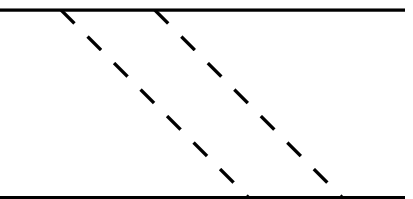}}
\label{sup1}
\end{eqnarray}

The last two graphs in (\ref{sup1}) containing two exchanged particles in the
intermediate state ("the stretched box") are omitted in the second iteration of
the light-front kernel. The number of  graphs with increasing number of
intermediate particles increases in higher orders. At small value of the
coupling constant $\alpha$ their contribution can be suppressed, but at
$\alpha\approx 1$ this reason of the suppression disappears.  {\em However,
these higher Fock state graphs are still suppressed}.

In the papers \cite{mariane,frederico} the binding energy was calculated in
the  framework of the Bethe-Salpeter equation and the light-front one. It was
found that even at $\alpha\approx 1$ the binding energies calculated in both
approaches are very close to each other. This indicates that the contribution
of the higher Fock states is suppressed.

\begin{figure}
\epsfxsize=8.5cm \epsffile{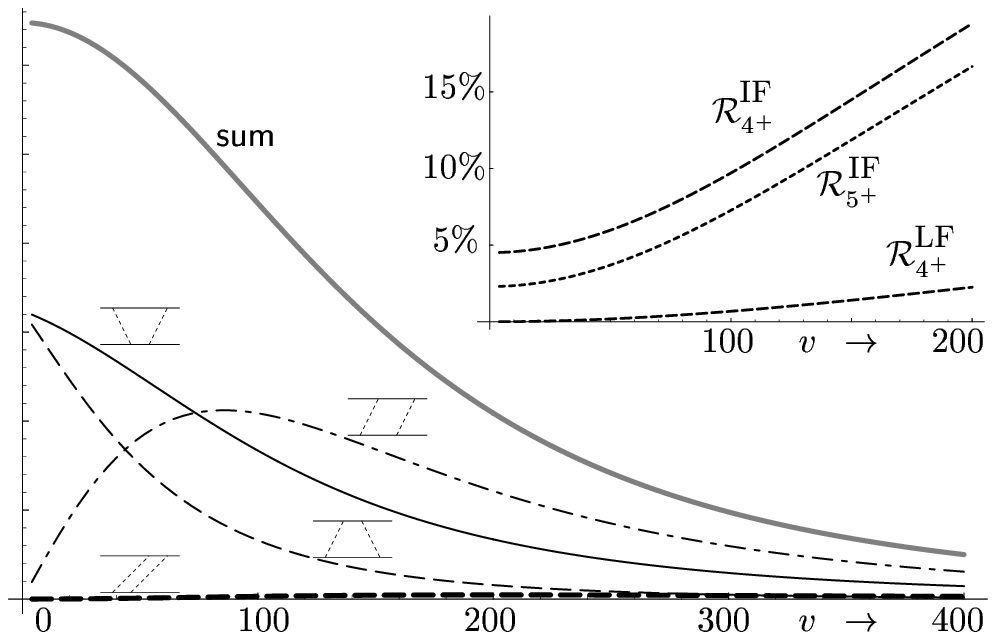}
\figcap{LF time-ordered boxes for a scattering angle
of $\pi/2$ as a function of the incoming momentum $v$. We also  
give the ratios of boxes with at least four particles 
(${\cal{R}}^{\rm IF}_{4^+}$ and ${\cal{R}}^{\rm LF}_{4^+}$) or five particles 
(${\cal{R}}^{\rm IF}_{5^+}$, ${\cal{R}}^{\rm LF}_{5^+} =0$) in one
of the intermediate states.}
\label{figR4and5}
\end{figure}

This contribution  has been calculated directly in the papers 
\cite{bak99,nico}. The result is shown in fig. \ref{figR4and5}. In these
figures $v$ means the incoming momentum. One can see that the contribution of
the  stretched box into the sum of time ordered graphs is neligible. Its
relative contribution ${\cal{R}}^{\rm LF}_{4^+}$ is of the order a few per
cent.

\begin{figure}
\hspace{1.5cm} \epsfxsize=5.5cm \epsffile{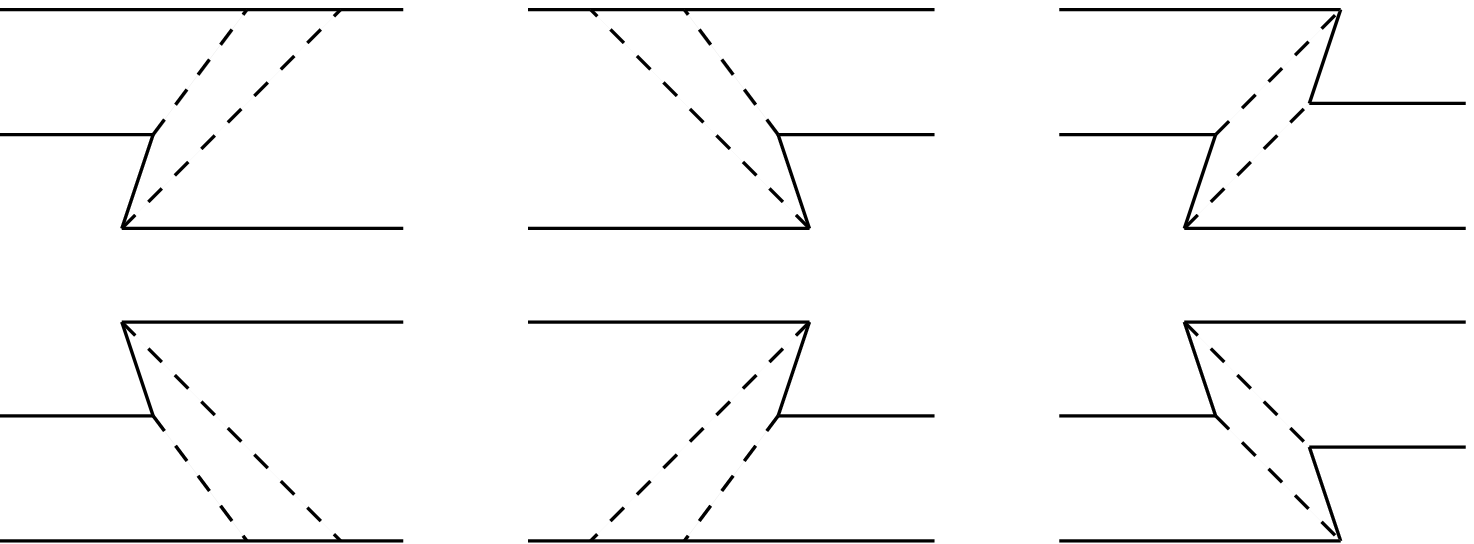}
\figcap{Time-ordered diagrams that contribute to ${\cal{R}}_5$. The diagrams in
the first column have five particles in the first intermediate state. The
diagrams in the second column have five particles in the last intermediate
state, and the diagrams on the right have five-particle intermediate states for
both the first and the third intermediate state.}
\label{figdiagsR5} 
\end{figure}

Another important conclusion which follows from fig. \ref{figR4and5} is that
the suppression of the higher Fock states takes place namely in LFD.  In the
instant form of dynamics these contributions much more larger.  For four or
more intermediate  particles,  due to the fluctuations, they are indicated in
fig. \ref{figR4and5} as ${\cal{R}}^{\rm IF}_{4^+}$. The corresponding graphs
are shown in fig. \ref{figdiagsR5}. For five or more intermediate  particles, 
due to a few vacuum vertices, they are indicated  as  ${\cal{R}}^{\rm
IF}_{5^+}$.

These results show that the light-front contributions of  higher Fock states
are significantly smaller than in the instant form.  In the limit  $v
\rightarrow 0$ the ratio ${\cal{R}}^{\rm LF}_{4^+}$ goes to zero, because the
phase space becomes empty. However, in the instant form there is a finite
contribution of ${\cal{R}}^{\rm IF}_{4^+} = 4.5 \%$ in this limit. 

%%%%%%%%%%%%%%%%%%%%%%%%%%%%%%%%%%%%%%%%%%%%%%%%%%%%%%%%%%%%%%

\section{Conclusion}\label{concl}

We have described the general construction of LFD, its explicitly covariant
formulation and some applications to the field theory and to the relativistic 
wave functions. These developments have been made particularly simple, and
intuitive, by the three-dimensional nature of formalism, interpretation of
amplitudes in terms of the space-time picture and the absence of vacuum
fluctuations. We have shown also the relation to other approaches, in
particular, to the Bethe-Salpeter one. 

Though the light-front amplitudes can be derived from the Feynman ones, and the
light-front wave function can be obtained by the projection of the
Bethe-Salpeter amplitude on the light-front plane, this does not mean that LFD
is only a method to calculate the  Feynman amplitudes and to find an
approximate eigenvalue of the Bethe-Salpeter equation. 

The light-front approach has much more general and independent meaning. It is
an alternative and rather powerful way to solve the field-theoretical problems.

%%%%%%%%%%%%%%%%%%%%%%%%%%%%%%%%%%%%%%%%%%%%%%%%%%%
\section{Appendix}\label{app}
\subsection{Kinematical transformations}\label{kt} 
We specify here the transformation properties of the state  vector with respect
to transformations of the coordinate system. 

The operators associated to the four-momentum and four-dimensional  angular
momentum are expressed in terms of integrals of the  energy-momentum
$T_{\mu\nu}$ and the angular momentum  $M^{\rho}_{\mu\nu}$ tensors over the
light-front plane $\omega\cd x  = \sigma$, according
to:                         
\begin{equation}\label{kt1}                                                     
P_{\mu}=\int T_{\mu\nu}\omega^{\nu}\delta(\omega\cd 
x-\sigma)d^4x=          
P^0_{\mu} +P^{int}_{\mu}\ ,                                         
\end{equation}                                                                  
\begin{equation}\label{kt2}                                                     
J_{\mu\nu}=\int M^{\rho}_{\mu\nu}\omega_{\rho} \delta(\omega\cd              
 x-\sigma)d^4x = J^0_{\mu\nu} +J^{int}_{\mu\nu}\ ,                 
\end{equation}                                                                  
where the $0$ and $int$ superscripts indicate the free and  interacting  parts
of the operators respectively. For generality, we consider here  the
light-front time $\sigma \neq 0$. 

The description of the evolution along the light-front time $\sigma$ implies  a
fixed value of the length of $\vec{\omega}$, or, equivalently, of  $\omega_0$.
This is necessary in order to have a scale of $\sigma$.  However, the most 
important properties of the physical amplitudes following from covariance do
not require to fix the scale of $\omega$ and will be invariant relative to its
change. We work in the interaction  representation in which the operators are
expressed in terms of the free  fields. Consider, for example, the scalar field
$\varphi(x)$, eq.(\ref{ft1}). Then the free  operators $P^0_{\mu}$ have the
form: 
\begin{eqnarray}\label{kt3}
 P^0_{\mu} &=&\int 
a^\dagger (\vec{k})a(\vec{k})k_{\mu}\ d^3k\ , \\ 
J^0_{\mu\nu}&=&\int                                                       
a^\dagger (\vec{k})a(\vec{k})i\left(k_{\mu}\frac{\partial}{\partial 
k^{\nu}}-          
k_{\nu}\frac{\partial}{\partial k^{\mu}}\right)d^3k\ ,\label{kt4}
\end{eqnarray}                                                                  
The operators $P^{int}$ and $J^{int}$ contain the interaction Hamiltonian 
$H^{int}(x)$:                                                       
\begin{eqnarray}\label{kt5}                                                                
P^{int}_{\mu}&=&\omega_{\mu}\int H^{int}(x)\delta(\omega\cd                  
x-\sigma)\ d^4x\ ,\\                                                
J^{int}_{\mu\nu}&=&\int H^{int}(x)(x_{\mu}\omega_{\nu} -x_{\nu}           
\omega_{\mu}) \delta(\omega\cd x-\sigma)\ d^4x\ .
\label{kt6}                      
\end{eqnarray}   
The field-theoretical  Hamiltonian  $H^{int}(x)$ is usually singular and
requires a regularization. The  regularization of  amplitudes  will be
illustrated above in sect. \ref{simple} by the example of a typical self-energy
contribution.

In the particular case $\omega=(1,0,0,-1)$, in the light-front coordinates,
only $\omega_-$-com\-po\-nent is non-zero.  This just gives that in
(\ref{kt4},\ref{kt6}) the components $P^{int}_-,J^{int}_{\perp,-}$ are
non-zero, i.e., corresponding generators in (\ref{kt1},\ref{kt2}) contains the
interaction. 

\begin{figure}[hbt]
\centerline{\epsfbox{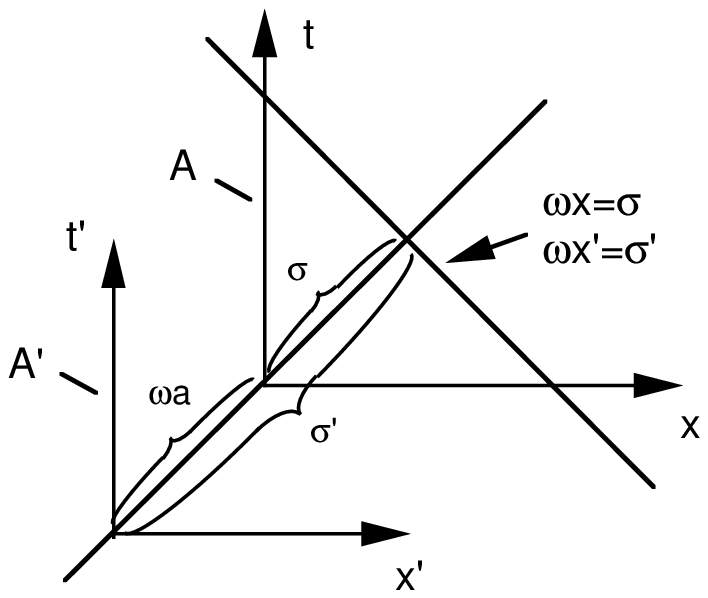}}
\figcap{Translation of the reference system along the 
light-front time.}
\label{refsys} 
\end{figure}

Under translation $x \rightarrow x'=x+a$ of the coordinate system 
$A\rightarrow A'$, the equation $\omega\cd x = \sigma$ takes the form  
$\omega\cd x'=\sigma'$, where $\sigma'=\sigma+\omega\cd a$. The state  vector
is transformed as:  
\begin{equation}\label{kt7}                                                     
\phi_\omega(\sigma)\rightarrow \phi'_\omega(\sigma ')                                         
  =U_{P^0}(a)\phi_\omega(\sigma)\ ,                                                    
\end{equation}                                                                  
where the operator $U_{P^0}(a)$ contains only the operator of the four-momentum
(\ref{kt3}) of the free field:  
\begin{equation}\label{kt8}
 U_{P^0}(a)=\exp(iP^0\cd a)\ .  
\end{equation}                                                                  
The ``prime" at $\phi'(\sigma)$ indicates that $\phi'(\sigma)$ is defined in
the system $A'$ on the plane $\omega \cd x'=\sigma$ in  contrast to
$\phi(\sigma)$ defined in the system $A$ on the plane  $\omega\cd x=\sigma$
(the value of $\sigma$ being the same). The state vector $\phi'(\sigma')$ is
defined in $A'$ on the plane $\omega\cd  x'=\sigma'$, which coincides with
$\omega\cd x = \sigma$. Therefore   no  dynamics  enters into the
transformation (\ref{kt7}). This is  rather  natural, since under translation
of the coordinate system the  plane $\omega\cd x = \sigma$ occupies the same
position in space while it occupies a new position with respect to the axes of
the new  coordinate system, as indicated in fig.~\ref{refsys}. The formal proof
of (\ref{kt7}), (\ref{kt8}) can be  found  in~\cite{karm82}.  

In the case of infinitesimal four-dimensional rotations $x_{\mu}  \rightarrow
x'_{\mu}=gx_{\mu}=x_{\mu}+\epsilon _{\nu\mu}x^{\nu}$, the  result is
similar~\cite{karm82}:                                                
\begin{equation}\label{kt9}                                                     
\phi_\omega(\sigma) \rightarrow 
\phi'_{\omega'}(\sigma)=U_{J^0}(g)\phi_\omega(\sigma)\ ,               
\end{equation}                                                                  
where $\omega'_{\mu}=\omega_{\mu} +\epsilon_{\nu\mu}  \omega^{\nu}$ and
\begin{equation}\label{kt9a}
U_{J^0}(g)=1+\frac{1}{2}J^0_{\mu\nu}\epsilon^{\mu\nu}.  
\end{equation} 
The operator $J^0_{\mu\nu}$ is given by (\ref{kt4}). This shows that the
transformations of the state vector with respect to the transformations of the
coordinate system are indeed kinematical.

%%%%%%%%%%%%%%%%%%%%%%%%%%%%%%%%%%%%%%%%%%%%%%%%%%%%%%%%%%%%%%%%%%%%% 

\subsection{Dynamical transformations}\label{dt} 
The properties of the state vector under transformations of the  hypersurface
are determined by the dynamics and follow from the  Tomonaga-Schwinger equation
\cite{qed}:  
\begin{equation}\label{kt10}                                                    
i\delta\phi/\delta\sigma(x)=H^{int}(x)\ \phi\ .                                 
\end{equation}                                                                  
From the definition of the variational derivative in (\ref{kt10}) we obtain:  
$$ 
i\delta\phi =H^{int}(x)\ \phi\ \delta V(x)\ ,                                   
$$                                                                              
where $\delta V(x)$ is the volume between the initial surface and the  surface
obtained from the original one by the variation  $\delta\sigma(x)$ around the
point $x$.  

Under the translation $\sigma\rightarrow \sigma +\delta\sigma$ of the plane,
the total increment of the state vector is obtained through the increment at
each point of the surface:                                     
\begin{equation}\label{kt11}                                                    
i\delta\phi =\int H^{int}(x)\delta(\omega\cd x -\sigma) d^4x\ \phi \ 
\delta       
\sigma\ .                                                                       
\end{equation}                                                                  
This relation gives the Schr\"odinger equation.  In the interaction             
representation in the light-front time, we have:                                 
\begin{equation}\label{kt11p}                                                                              
i\partial\phi/\partial\sigma = H(\sigma)\phi(\sigma)\ ,                         
\end{equation}                                                                               
where:                                                                          
\begin{equation}                                                                              
H(\sigma) =\int H^{int}_{\omega}(x) \delta(\omega\cd x-\sigma)d^4x,            
\end{equation}  
and $H^{int}_{\omega}(x)$ may differ from $H^{int}(x)$ because of 
singularities of the field commutators on the light cone. This point  is
explained below in the section \ref{smat}.

Similarly, in the case of a rotation of the light-front plane, 
$\omega_{\mu}\rightarrow \omega'_{\mu}=\omega_{\mu}+ \delta\omega_{\mu},$
$\delta\omega_{\mu}=\epsilon_{\nu\mu}\omega^{\nu}$, we find:
\begin{equation}                                                                                
\phi_\omega(\sigma) \rightarrow \phi_{\omega+\delta\omega}(\sigma) =\phi_\omega                       
+\delta\phi_\omega,\; \delta\phi_\omega=\frac{1}{2}\epsilon_{\mu\nu}                          
\left( \omega^{\mu}\frac{\partial}{\partial\omega_{\nu}} 
-\omega^{\nu}          
\frac{\partial}{\partial\omega_{\mu}}\right)\phi_\omega(\sigma)\ .                     
\end{equation}                                                                               
The increment of the volume over the point $x$ is:                              
\begin{equation}                                                                               
\delta V =\epsilon_{\mu\nu}\ x^{\mu}\ \omega^{\nu}\ \delta(\omega\cd               
x-\sigma)\ d^4x\ ,                                                             
\end{equation}                                                                               
and it follows from (\ref{kt11}) that~\cite{karm82}:                            
\begin{equation}\label{kt12}                                                    
J^{int}_{\mu\nu}            
\ \phi_\omega(\sigma)= L_{\mu\nu}(\omega)\phi_\omega(\sigma)\ ,                                                               
\end{equation}                                                                  
where:                                                                          
\begin{equation}\label{kt13}                                                    
L_{\mu\nu}(\omega) =i\left(\omega_{\mu}                                   
\frac{\partial}{\partial\omega^{\nu}} -\omega_{\nu}                             
\frac{\partial}{\partial\omega^{\mu}}\right)\ ,                                 
\end{equation}                                                                  
and $J^{int}_{\mu\nu}$ is given by (\ref{kt6}).                           
                                                                                
Equation (\ref{kt12}) is called the {\it angular condition}. It plays  an
important role in the construction of relativistic bound states.

The transformation of the coordinate system and the simultaneous 
transformation of the light-front plane, which is rigidly related to the 
coordinate axes, correspond to the successive application of the two types of 
transformations considered above (kinematical and dynamical). Thus, under the
infinitesimal  translation $x\rightarrow x'=x+a$ of the coordinate system,
$A\rightarrow A'$, and of the plane, we have:                                
\begin{equation}\label{kt13a}                                                   
\phi_\omega(\sigma)\rightarrow \phi'_\omega(\sigma)=(1+iP\cd 
a)\phi_\omega(\sigma)\ .              
\end{equation}                                                                  
Note that for the state with definite total four-momentum $p$ (i.e., for an
eigenstate of the four-momentum operator), the equations (\ref{kt7}) and
(\ref{kt13a}) give:  
\begin{equation}\label{kt18a}                                                   
\exp(iP^0\cd a)\phi(\sigma) =\exp(ip\cd a)\phi(\sigma 
+\omega\cd a)\ .             
\end{equation}
This equation determines the conservation law (\ref{sc1}) for the four-momenta
of the constituents.

\end{document}